\documentclass[times,twocolumn,5p]{elsarticle}

\usepackage{hyperref}
\usepackage{subfig}
\usepackage{booktabs}
\usepackage{tabularx}
\usepackage{eurosym}
\usepackage[english]{babel} 
\usepackage[shadow,backgroundcolor=green!40]{todonotes}
\usepackage{multirow}

\newcolumntype{Y}{>{\centering\arraybackslash}X}
\journal{Future Generation Computer Systems}

\begin{document}

\begin{frontmatter}

\title{Runtime Verification for Business Processes Utilizing the Bitcoin Blockchain}

\author[tuv]{Christoph Prybila}
\ead{c.prybila@infosys.tuwien.ac.at}
\author[tuv]{Stefan Schulte\corref{cor1}}
\ead{s.schulte@infosys.tuwien.ac.at}
\author[tuv]{Christoph Hochreiner}
\ead{c.hochreiner@infosys.tuwien.ac.at}
\author[data]{Ingo Weber}
\ead{ingo.weber@data61.csiro.au}

\address[tuv]{TU Wien, Vienna, Austria}
\address[data]{Data 61, Sydney, Australia}
\cortext[cor1]{Corresponding author}

\begin{abstract}
The usage of process choreographies and decentralized Business Process Management Systems has been named as an alternative to centralized business process orchestration. In choreographies, control over a process instance is shared between independent parties, and no party has full control or knowledge during process runtime. Nevertheless, it is necessary to monitor and verify process instances during runtime for purposes of documentation, accounting, or compensation. 

To achieve business process runtime verification, this work explores the suitability of the Bitcoin blockchain to create a novel solution for choreographies. The resulting approach is realized in a fully-functional software prototype. This software solution is evaluated in a qualitative comparison. Findings show that our blockchain-based approach enables a seamless execution monitoring and verification of choreographies, while at the same time preserving anonymity and independence of the process participants. 
Furthermore, the prototype is evaluated in a performance analysis.
\end{abstract}

\begin{keyword}
Choreographies, Blockchain, Business Process Management, Runtime Verification

\fbox{\parbox{2\linewidth}{
		NOTICE: This is the authors' version of the accepted manuscript published in Future Generation Computer Systems. Please cite as: \textbf{Christoph Prybila, Stefan Schulte, Christoph Hochreiner, Ingo Weber: Runtime Verification for Business Processes Utilizing the Bitcoin Blockchain. Future Generation Computer Systems}. The  version layouted by Elsevier can be found using the following DOI: \url{https://doi.org/10.1016/j.future.2017.08.024}. The content of this paper and the one available at \url{https://doi.org/10.1016/j.future.2017.08.024} is identical.
		\begin{flushleft}
					\copyright 2017. This manuscript version is made available under the CC-BY-NC-ND 4.0 license \url{http://creativecommons.org/licenses/by-nc-nd/4.0/}
		\end{flushleft}
}}
\end{keyword}
\end{frontmatter}

\section{Introduction}
\label{sec:intro}
Business Process Management (BPM) comprises methods and concepts to model, execute, monitor, configure, and administrate the processes which are at the core of a company's business. These processes are made up from a set of \textit{activities} (or \emph{tasks}) which together fulfill a business goal~\cite{weske12}. The composition and control of such a set of activities can either be done in a centralized way (\emph{orchestration}), where the process owner has full control and knowledge about all tasks at all points of time, or in a decentralized way (\emph{choreography}), where process control is handed over from the process owner to other process participants and no partner has full control and knowledge of the process~\cite{mendling13}. In a choreography, each involved participant receives information about the negotiated terms and requirements of a designated part of the choreography and then acts independently \cite{leite12}.

Choreographies are especially important for inter-organizational settings, where it is not possible to establish a centralized process control because of organizational boundaries \cite{mendling13}. Furthermore, in very large settings, a central Business Process Management System (BPMS) may become a bottleneck \cite{schulte15}. Choreographies, which are executed in a highly decentralized manner, help to mitigate this scalability issue \cite{leite12}.

Especially in \emph{Business to Business} (B2B) settings, contracts are at the core of processes~\cite{norta15}. A process contract describes the terms of the collaboration between different entities, i.e., the obligations and rights of all process participants, and the rules applying to process execution. Such a contract is attached to a process instance. In the case of choreographies, records about the decentralized execution of a process are the basis for contract verification. This documentation must be indisputable and accepted by all choreography participants~\cite{baouab11}. It can then be used to enforce the contract underlying a choreography. Based on this, a process owner can penalize a process participant in case of the incorrect execution of an activity. At the same time, a participant can claim payment from the process owner for the participation in the execution of a process, since the execution of an activity is documented. 

To enable this kind of documentation, a choreography-oriented BPMS must incorporate an end-to-end process verification mechanism. Verification is defined as the ``evaluation of whether or not a product, service, or system complies with a regulation, requirement, specification, or imposed condition''~\cite{pmbok13}. With regard to business processes, we define runtime verification as the evaluation whether a process execution meets the functional and non-functional objectives defined in a contract between the process participants. A feature like this enhances the overall trust into the robustness of choreographies and the acceptance of the overall process-based cooperation between the involved participants. 

Cryptocurrencies like Bitcoin document and verify conducted payments in a decentralized ledger, called \emph{blockchain}~\cite{zohar15}. Through cryptographic security measures, the funds of each single party are protected. During the process of paying another party, both parties must have undeniable proof that the correct amount of money was indeed sent. Blockchains are not maintained by a single financial institution but by a large number of small and independent peers, called \emph{miners}~\cite{alqassem14}. This increases the trust in a blockchain as an independent institution.

Both choreography-oriented BPM and cryptocurrencies face similar challenges when performing verification. The parties involved in cryptocurrency payment transfers are independent and mostly even anonymous. Nevertheless, payments which have been conducted must be permanent and indisputable~\cite{tschorsch16}. Choreography participants are also independent, and in some cases even potential competitors~\cite{vonriegen09}. Hence, the performed tasks of a choreography process instance must also be permanently documented in a trusted way. For this, the utilization of a blockchain for choreography-oriented BPM appears to be a promising approach. The goal of the work presented in this paper is therefore to determine the suitability of blockchains to serve as trust basis for decentralized and indisputable runtime verification for choreography-oriented BPM. 

Our contributions can be summarized as follows:
\begin{itemize}
	\item We examine the usage of the blockchain technology as means to establish trust and to allow verification of choreographies. For this, we discuss different runtime verification approaches and briefly evaluate the properties of existing blockchains regarding their suitability for this task.
	\item We develop a blockchain-based runtime verification approach and implement a fully-functional prototype which is able to verify running process instances using the Bitcoin blockchain. 
	\item We evaluate the capabilities of the developed prototype in a qualitative comparison and a performance analysis.
\end{itemize}
The remainder of this paper is structured as follows: In Section~\ref{sec:background}, we present background work on BPM and blockchains and define prerequisites for our work. Afterwards, we discuss the related work in Section~\ref{sec:related}. In Section~\ref{sec:approach}, we introduce our approach to runtime verification in choreographies. We evaluate the presented approach in Section~\ref{sec:evaluation}. Section~\ref{sec:conclusion} concludes this paper and provides a short outlook on our future work.

\newpage
\section{Background}
\label{sec:background}
The following subsections present background information regarding the technical content of this paper. For this, we discuss BPM (Section~\ref{sub:bpm}) and blockchains (Section~\ref{sub:blockchains}). Section~\ref{subsec:basicBitcoinTransactions} describes the basic structure of Bitcoin transactions. Afterwards, we define prerequisites for our work (Section~\ref{sub:prerequisites}). 
\subsection{Business Process Management}
\label{sub:bpm}
\begin{figure}[t]
	\centering
	\includegraphics[width=\columnwidth]{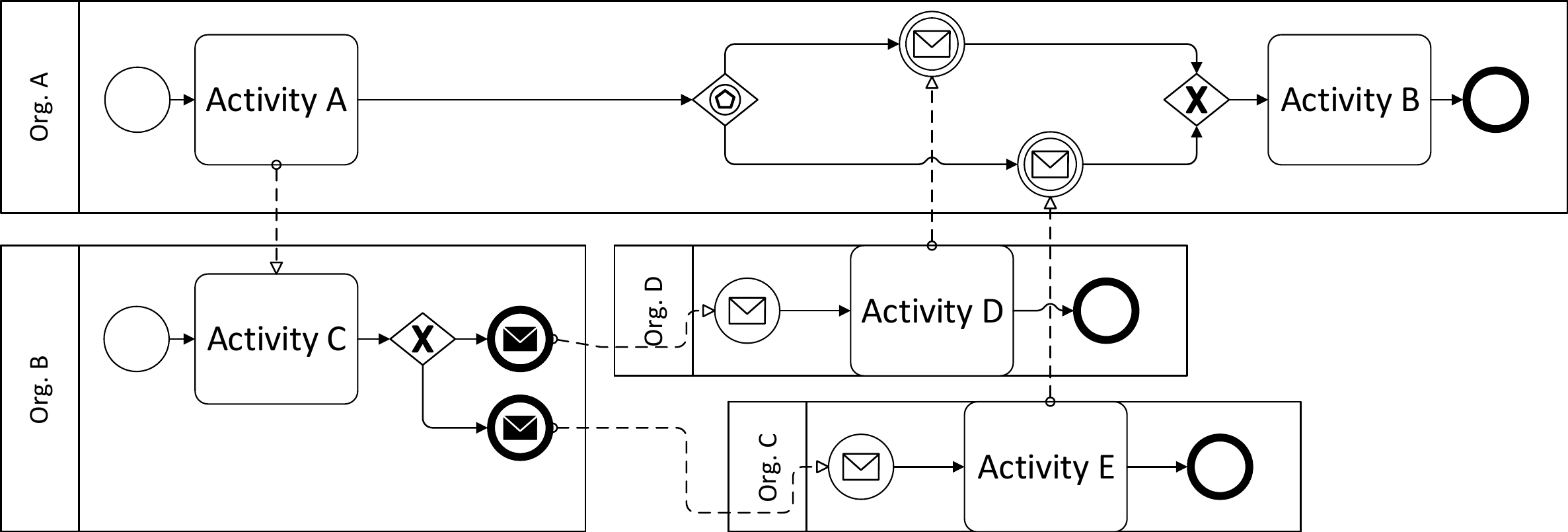}
	\caption{Example Choreography, adapted from \cite{mendling13}}
	\label{fig:choreography}
\end{figure}
Classic approaches to BPM are usually based on a centralized BPMS, which orchestrates business processes~\cite{mendling13}. Such a BPMS becomes the single contact point for all process instances \cite{tsai10}. More recent approaches propose a different solution where multiple services communicate directly with each other in a peer-to-peer (P2P) fashion and the responsibility for process control is shared between different entities, i.e., processes are choreographed \cite{eder08, ahmed14}. In such settings, there is no need for a centralized BPMS.

\textit{Choreographies} distribute the control of executed process instances to different independent process participants. Each participant receives information about the negotiated terms and requirements of a designated part of the choreography and then acts independently. Messages are exchanged directly in a P2P fashion between the involved participants. This design removes single points of failure as well as messaging bottlenecks~\cite{leite12}. Choreography-oriented BPM especially benefits use cases involving B2B cooperation \cite{weske12}, where inter-organizational processes need to be executed. In such settings, companies often hesitate to open their systems to partners~\cite{norta15,wetzstein10}, which would be needed to orchestrate a process and have global knowledge about a process' status. An example choreography is illustrated in Figure~\ref{fig:choreography}. 

The choreography approach creates several opportunities compared to centralized orchestrations. By distributing control and execution, scalability and robustness are improved~\cite{tsai10}. Also, choreographies are a promising solution for inter-organizational processes to bridge information gaps caused by organizational borders between process participants. 

At the same time, choreographies create new challenges. Decentralized execution requires process participants to hand over the control of process instances to remote partners. The formed cooperation contracts need to be verifiable to make them enforceable for accounting and compensation purposes. Furthermore, process owners must be able to trace the execution path of a process instance across the boundaries of the different process participants. At the same time, a participant has to be able to proof its participation in a choreography. The collected information must be trustworthy to serve as legal basis for contract enforcement.

Today, a dedicated runtime verification approach for choreographies, which fulfills the requirements stated above, is missing~\cite{anseeuw15,montagut08} (see Section~\ref{sec:related}). Hence, new solutions need to be found. As outlined in Section~\ref{sec:intro}, we therefore propose a blockchain-based approach to enable runtime verification for choreographies.

\subsection{Blockchain}
\label{sub:blockchains}

The original blockchain concept stems from the cryptocurrency Bitcoin~\cite{nakamoto08}, which aims to provide a decentralized mechanism for the anonymous exchange of digital money. One particular issue with regard to digital money is the \emph{double spending problem}, i.e., a malicious or faulty party could duplicate digital money tokens and spend them multiple times \cite{tschorsch16}. Therefore, it is necessary that the amount of money one party holds is recorded in a ledger. A digital payment transfer can then be conducted by reducing the balance in one party's ledger while increasing the balance of another party's ledger. 

These ledgers must be managed by a trusted institution. With regard to cryptocurrencies' goal to achieve total anonymity of payers and payees, there should be no central party who controls the balances of ledgers and the transfer of money between two particular parties. Hence, in cryptocurrencies like Bitcoin, book-keeping and verification of payment transfers is not controlled by one particular party (or a group of parties), but by a vast number of independent parties, the so-called \emph{miners}, which together operate a blockchain~\cite{zohar15}. In the following paragraphs, the basic technical approach of the Bitcoin blockchain is discussed. Other well-known first generation blockchains, e.g., the Litecoin blockchain, are based on the same basic principles, while so-called second generation blockchains extend the possible application areas (see below)~\cite{tschorsch16}.

A blockchain is represented by a chain of transactions \cite{nakamoto08}. In Bitcoin, these transactions are public~\cite{zohar15}. Hence, everyone can fetch the historical transaction data and determine how many funds are associated with certain Bitcoin addresses and what payment transfers have been conducted. Anonymity is achieved through the abstraction of Bitcoin addresses. 

When a transaction is broadcast to the Bitcoin network, it is first verified and then added into a new block. New blocks are placed on top of the previous transaction history, i.e., the blockchain. As a result, the blockchain provides an unchangeable history of all Bitcoin transactions. As the name points out, the blockchain consists of a series of interconnected data blocks. Each block contains a number of transactions as well as a link to the previous block, incentive information, and a \emph{proof of work}~\cite{alqassem14}. 

Importantly for some design decisions made in the work at hand, blockchains are categorized into first generation and second generation blockchains \cite{weber16}. The first operating blockchains were designed to serve a single main purpose and have limited adaptability for other use cases. They are referred to as first generation blockchains, e.g., Bitcoin or Litecoin. 

Recent implementations of second generation blockchains like Ethereum do not intentionally limit the use cases that can be addressed. 
By providing an open programming environment in their blockchains, they aim to support and facilitate various use cases. 
Hence, Ethereum is not limited to supporting a cryptocurrency and enables so-called \emph{smart contracts}. These contracts are program agents that are executed within the Ethereum environment. They are defined in a Turing-complete scripting language and enriched with private storage as well as monetary balance.  
During its execution, a contract is able to interact with its storage, send messages, and create other contracts.

On the one hand, the open nature of Ethereum is certainly suited to support new prototypes, but on the other hand, the Ethereum blockchain currently suffers from major stability issues. In contrast, the Bitcoin blockchain on the other hand has proven to be the most mature and robust framework among all blockchains. Our runtime verification prototype is therefore based on the Bitcoin blockchain. An in-depth discussion about the blockchain selection can be found in Section \ref{sec:related}.

\subsection{Basic Bitcoin Transactions}
\label{subsec:basicBitcoinTransactions}

Bitcoin transactions are tailored to conduct payments. A common transaction is composed of an input section and an output section. The \emph{owner} of a Bitcoin has access to the output of the latest transaction in which the corresponding Bitcoin was used. To spend the Bitcoin, the owner has to issue a transaction in which the output of the previously latest transaction becomes the input to a new transaction \cite{tschorsch16}. The payer (i.e., the transaction sender) specifies the new owner of the Bitcoin by directing the output of the new transaction to a specific Bitcoin address. Since the output of the previously latest transaction now has been used, it is considered spent and cannot be used as input for another transaction. After creating the overall transaction information, the data is signed by the private key of the payer's Bitcoin address and broadcast to the Bitcoin network to become part of the blockchain~\cite{donet14}.

A standard Bitcoin transaction can have multiple inputs and outputs~\cite{zohar15}. Thus, the input of a single new transaction can be composed of multiple parts referencing the outputs of multiple old transactions. Likewise, the resulting amount of a transaction can be split into multiple output parts. These output parts can then be distributed to multiple Bitcoin addresses. This enables the payer to pay multiple parties at once and to receive change at the same time. 

The technical mechanics of a standard transaction are as follows: Output parts are sums of Bitcoins paired with small scripts written in a custom scripting language. These scripts guard the funds associated with the output. To access an output part, the corresponding script must be supplied with a parameter that renders its result to \emph{true} \cite{zohar15}. The structure of a Bitcoin transaction is well-defined. Most fields expect specific content and have fixed lengths. The included scripts are the only exception. They do not have fixed field lengths and their content can in theory be defined arbitrarily.

Nevertheless, most scripts follow certain patterns. The most common script requires a signature as input, created from the payee's Bitcoin private key. Only the holder of the corresponding private key is able to create the required signature, thus making the output only accessible to the owner of the destined Bitcoin address~\cite{alqassem14}. The input part of a transaction only contains a reference to its corresponding originating output part and the necessary parameters to render the script of the output to \emph{true}. Through this mechanism, everybody receiving a broadcast transaction can verify if the transaction is really authorized to access the specified outputs.

While the Bitcoin blockchain was established for a specific use case, its original design provides various possibilities to create experimental transactions. We will exploit these possibilities for our process runtime verification approach presented in Section~\ref{sub:approach}.

\subsection{Prerequisites}
\label{sub:prerequisites}
The following paragraphs define some prerequisites and settings for our work. 

Despite the lack of centralized control in a choreography, we assume that there is always a process owner, who initiates a business process, shares the process model with all participants, and is paying for its successful decentralized execution~\cite{norta15}. The functional and non-functional execution constraints and monetary rewards of a process activity are defined by the process owner, in many cases described as Service Level Agreements (SLAs), and realized in a contract between process owner and process participants. This bundled process information is shared among all choreography partners. 

With regard to choreographies, we assume that the involved parties are reluctant to share information about their identities, data, or internal business structures~\cite{baouab11,vonriegen09}. However, the process owner requires information about the executed decentralized process activities. Most importantly, the process owner needs to know which activities have been fulfilled by which partners and how long the execution took~\cite{montagut08,weber16}. For this, especially the handover of the control of a process instance must be documented in an undeniable way and must be accessible for the process owner. Since decentralization is inherent to choreographies, there is no centralized control and knowledge about process instances. 

We assume that the selection of choreography participants is highly flexible and allows the selection of partners during process runtime. This enables, e.g., to replace one process participant (respectively the activity to be executed by the participant) during the runtime of a process. 

For this, certain assumptions are made. It is assumed that the process owner initially hands over the execution of the process to a suitable partner to have a specific process activity executed. To accomplish this, the process owner first selects the next suitable choreography participant. This participant is either predefined or chosen based on the required service and the defined SLA. If selected ad hoc, the process owner and the selected potential cooperation partner negotiate the terms of the handover. These selection and negotiation steps are well-covered in Service-oriented Architectures (SOA), e.g., ~\cite{benbernou10}, and will therefore not be discussed in-depth in this paper. After a process participant has finished the execution of the defined task, the control over the process execution is passed along to the next choreography participant. This is done by employing the previously described selection and negotiation steps. 

Notably, all process participants remain independent organizations and are potential competitors~\cite{vonriegen09}. That is why one goal for runtime verification is to keep mutual dependencies to a minimum. 
Nevertheless, the handover of a process instance to another participant together with the achieved progress must be documented. For this, no centralized invasive monitoring service can be used, since such monitoring would introduce a tighter coupling and increased requirement for information sharing between the participants of a choreography.

Finally, we do not limit the kinds of processes or tasks supported by our runtime verification approach. Tasks can be software-based services or real-world activities. As long as it is possible to monitor the handover between the process participants that offer the respective tasks (e.g., by a decentralized BPMS), our approach is able to verify process executions.

\section{Related Work}
\label{sec:related}
The monitoring of choreographies has been investigated from different angles. Norta et al. argue that as a foundation for choreography monitoring, it is first necessary to negotiate a contract between the parties involved~\cite{norta15}. For this, a markup language is provided which supports the exchange of choreography contracts. This language allows to define what monitoring information should be provided by which choreography participant and how this information is shared and accessed. 

In addition to contractual definitions, monitoring in decentralized process execution needs to be defined during choreography modeling. Ansseuw et al. provide an extension to the standard BPMN~2.0 monitoring injection points to support choreographies \cite{anseeuw15}. Wetzstein et al. also propose a technology-driven approach to enable choreography monitoring by extending BPEL4Chor with an event-oriented monitoring agreement~\cite{wetzstein10}. To address privacy concerns of process participants, events can only be defined based on a publicly available process model. How each participant maps public choreography activities to internal processes remains hidden. None of the approaches discussed so far makes use of a decentralized ledger to store and distribute monitoring data in a trustworthy way. Notably, if this is supported at all, it is necessary to explicitly define which data is shared with whom. 

With regard to runtime verification of choreography executions, two general approaches have been proposed: von Riegen and Ritter \cite{vonriegen09} propose the usage of multiple \emph{Enterprise Service Buses} (ESBs) to handle all communication between the cooperating parties. The authors suggest the usage of proxies which intercept all communication and log all necessary information in a central component. 
A similar approach is described by Baouab et al. \cite{baouab11}. In their scenario, the cooperating participants of a choreography are already chosen at deployment time by the process owner. 
To guarantee the enforcement of given policies, all participants must run the same communication gateways which intercept all traffic. If any deviations are observed, events are emitted to notify the process owner.

The second general approach to runtime verification of choreography executions proposes \textit{token passing} along the process participants. By enhancing the token with cryptography features, it becomes a proof for the path it traveled along. Through keeping a copy, each partner can proof its participation in the corresponding process instance. Upon receiving the corresponding tokens, process owners are able to verify the exact sequence of execution. Depending on the structure of a choreography's required data sources, this technique can also be used to ensure data integrity and confidentiality. If the required data can be sent along the choreography as a single document, the document becomes the token upon which the security features are applied. Examples for this approach have been presented, e.g., in \cite{montagut08, bengtsson05, lim12}. 

Message controlling and token passing are two approaches for controlling and propagating the execution state of a choreography. Both impose different challenges when being applied to the scenario regarded in this paper. If loose coupling is a priority in the cooperation environments, the runtime verification approach of message controlling becomes difficult. Communication frameworks like message buses can ensure that choreography messages passed between the participants only use the provided connectors. But to set up a such communication framework between partners, a tight integration between the software systems becomes necessary, which also leads to a high rigidity of the choreographies that can be supported by these approaches. The token-based approach also introduces new problems, since the controlling token might get lost. While in theory in our approach also a private key could get lost, these keys should not be shared between participants. Also, each participant creates a unique private key for each transaction. 

To avoid token loss, choreography runtime implementations for BPMS like the ones presented by Martin et al. \cite{martin08} and Hwang et al. \cite{hwang13} save the tokens in a shared storage. This shared storage then becomes the controlling entity for the system. The shared storage must be operated by a trusted third party, which does not fit the basic assumption of no centralized source of trust in choreography-based BPM.

As described in the previous sections, we solved the problems of these approaches by using the Bitcoin blockchain as the trusted entity for the choreography. In general, the usage of blockchain in BPM is still at its very beginning~\cite{MWA+17}. Nevertheless, through its design, the blockchain can provide a shared trust basis which is not under the control of a single organization. Messages can be exchanged directly within blockchain transactions and token information can be stored in the blockchain by embedding them in transactions. 

In parallel to the work at hand, Weber et~al. proposed a related approach. In this solution, Ethereum smart contracts are used for runtime verification of choreographies~\cite{weber16}. For this, the authors present a component which translates BPMN models into smart contracts. Instantiated contracts become the controlling entities of process instances and are able to control and document all process activities. During the creation of an instance contract, the public keys of the choreography's participants together with their corresponding roles must be provided. Partners do not communicate directly with each other. Instead, they interact through transactions which are submitted against an instance contract and its contract storage. These transactions alter the state of the given contract and at the same time advance the execution state of the given process instance. Since all transactions are verified against the contract definition, it can be ensured that only authorized participants can alter the process state at a given execution point. Furthermore, the execution sequence can be enforced to match the process definition. At last, all this information is publicly documented in the Ethereum blockchain.
Building on the initial paper, Garcia et al.~\cite{Garcia2017} present an improved version of the approach by Weber et al.~\cite{weber16} in terms of cost efficiency, throughput, and latency.

The main downside of the seminal approach by Weber~et~al. is that all participants of the choreography must be known in advance. This is required to include the relevant public keys and roles into a contract during its creation. This reduces the flexibility of the overall choreography and makes the process execution less robust. If one of the participants is unreliable or becomes unreachable, the whole process execution may be stuck. Without the public key and role information, access security and execution sequence enforcement are not possible. 

While the approach proposed in this paper is based on the Bitcoin blockchain, Weber et al. utilize the Ethereum blockchain. The Ethereum blockchain provides great programmatic freedom, however, there is a lack of systematic analyses of Ethereum with regard to its stability and general quality, e.g., in terms of how the miners are distributed and if they are independent from each other. The Ethereum blockchain has been subject to heavy attacks in 2016, leading to several \emph{hard forks} to restabilize the blockchain and make it more secure.\footnote{\url{https://blog.ethereum.org/2016/10/18/faq-upcoming-ethereum-hard-fork/}, accessed on 31.10.2016\\\url{https://blog.ethereum.org/2016/11/18/hard-fork-no-4-spurious-dragon/}, accessed on 30.11.2016} Because of these instabilities, it is currently not advisable to apply the Ethereum blockchain in a B2B context, where stability and trust are of uttermost importance.

In contrast, there have been quantitative and qualitative analyses of the Bitcoin network~\cite{donet14, yeow16}. In these analyses, the authors have shown that the Bitcoin network is distributed worldwide and that approximately 6,000 nodes form its reliable core~\cite{donet14}. Since the Ethereum blockchain is less mature and less stable than the Bitcoin blockchain, we decided to utilize the latter in the B2B scenario applied in the work at hand.

One particular issue of Bitcoin is however its limited scalability. The Bitcoin project struggels to provide the transaction throughput required for the current demand. To increase the blockchain throughput, either the block size or the block creation frequency must be increased. Both factors influence the network's capability to synchronize in time~\cite{decker13}. This would result in an increase in conflicting blocks~\cite{donet14} and a reduced security level. While different solutions for this issue have been researched, e.g.,~\cite{sompolinsky15}, none has been integrated yet into Bitcoin.

\section{Runtime Verification for Choreographies}
\label{sec:approach}
As described in Section~\ref{sec:background}, the Bitcoin blockchain offers a promising basis to implement independent, decentralized, and undeniable runtime verification for choreographies. This section analyses this approach in more detail. In Section~\ref{sub:approach}, we present our conceptual approach to runtime verification. In Section~\ref{sub:implementation}, we describe how we realize the conceptual solution with Bitcoin transactions, and in Section~\ref{sub:framework}, a framework which encapsulates the implementation is presented.

\subsection{Conceptual Solution}
\label{sub:approach}
\subsubsection{Basic Approach}
Because of our decision to make use of the Bitcoin blockchain, only transaction techniques of existing first generation blockchains are used, instead of utilizing smart contracts of second generation blockchains like Ethereum. This limits the possible features but enables the usage of existing well-supported and mature blockchains, like Bitcoin. On the downside, the runtime verification proposal has to address the limited adaptability (in terms of supported use cases by the original protocol) and scalability (see Section~\ref{sec:related}) when using Bitcoin.

In our approach, a free Bitcoin output is selected by the process owner at the start of a new process instance to serve as the control token for the choreography. This control token stores the execution state of the process. At the same time, the blockchain becomes the decentralized storage for the token. Whoever is in possession of the token is responsible for the execution of a part of the choreography, i.e., an activity/task. To enable parallelism in a process instance, the token can be split and joined. Participants can document the progress of the process and of the handover to other participants by submitting new transactions which propagate the token. 

Each transaction is enriched with additional metadata about the current state of the process. Since Bitcoin transactions are push-based, the token sender gives its approval of a handover from one participant to another by publishing the respective transaction. Nevertheless, also the approval of the token receiver must be documented in the transaction. Therefore, a signature of the token receiver is also embedded in the process metadata stored in the transaction. 

The transaction chain related to the token of a process instance provides undeniable proof about the process' progress. If this progress somehow violates the agreements of the choreography contract, penalties can be claimed by the process owner from the involved participants. At the same time, it is possible for participants to prove their successful involvement in a choreography to claim their rewards. To preserve the flexibility of the choreography, the participants are not predetermined at process design time, but can be chosen dynamically during process runtime. On the downside, this prevents the enforcement of a correct process sequence. Still, it is not possible for a single participant to forge critical blockchain entries. Therefore, a process owner can monitor the progress of an instance by observing the blockchain. If the execution of a certain process instance deviates from the given process model, a process owner and all other choreography participants of this instance can detect this and react to it.

\subsubsection{Token Information}
To change a Bitcoin transaction into a documentation element which proves that a process instance has been handed over from one participant to another, it must provide the following characteristics: First, token handovers must be access-protected. Only the current owner of a process token must be able to decide whom to pass on the token. Since the token is essentially an amount of Bitcoins, this kind of access protection is already a built-in feature of Bitcoin. Each output of a Bitcoin transaction is protected by a script which commonly requires a Bitcoin signature of the owner as parameter.
At the same time, the receiver of the token must confirm that a handover, together with the included metadata, is accepted. 
When the handover documentation element is completed, signatures of both sender and receiver must be contained. 

The following information has to be included in the transaction to document the state of the current process execution:

\begin{description}
	\item[process id:] To capture which instance is addressed by the transaction, the identification number of the process must be included. 
	\item[next task id:] A participant works on a specific process task (i.e., activity) and then hands over the control of the process to another participant to perform the next task. The identification number of the task which should be performed by the receiving participant must be included in the transaction.
	\item[timestamp:] This timestamp documents the point in time the current task, processed by the sending participant, ends and the following task, processed by the receiving participant, starts. This is important to assess if deadlines have been met.
	\item[process data hash:] Most process instances require data to operate on. This data is continuously altered by the fulfilled process tasks. To document the current state of the process data, a hash must be placed in the transaction. The actual data transfer takes place off-chain, directly between the involved participants.
	\item[receiver signature] Not only the sender must confirm the handover of a process, also the agreement of the receiver must be documented. Therefore, the receiver must also sign the transaction template before publishing. In this transaction template, all the data described above must already be included. This way, the receiver documents the approval to receive control over the process instance under the documented conditions.
\end{description}
Finally, identification data of sender and receiver must be exchanged. By design, Bitcoin transactions are sent between Bitcoin addresses. For the work at hand, new addresses are generated for each handover. These addresses are anonymous and protect the privacy of the involved choreography participants. 

Still, sender and receiver must be able to mutually prove their handover. 
Therefore, it is assumed that besides the Bitcoin infrastructure, a RSA-based \emph{public key infrastructure} (PKI) is in place. By utilizing RSA-based signatures and certificates, a participant can prove its identity to others~\cite{cormen09}. When sender and receiver want to perform a handover, they first have to share the respective Bitcoin addresses they want to use. This exchange is enriched with RSA-based signatures and certificates. This way, each handover partner confirms the ownership to a given Bitcoin address before the handover takes place. By storing the received signature, a choreography participant can prove the identity of the corresponding handover partner to the process owner, if required. 

\subsubsection{Handover}
\label{subsub:handoverprocess}
\begin{figure}[t!]
	\centering
	\includegraphics[width=\columnwidth]{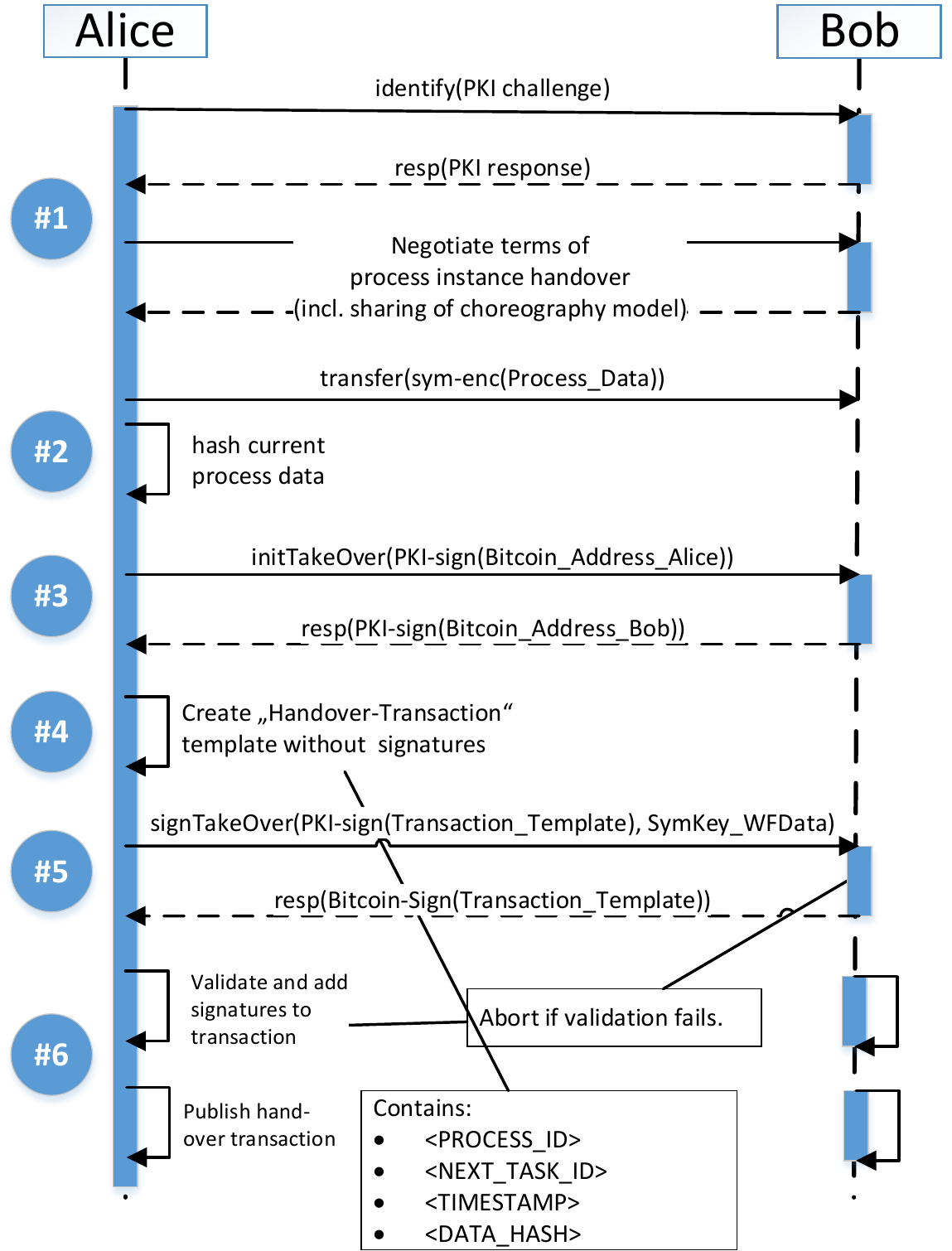}
	\caption{Intermediate Handover of a Process Instance between Choreography Participants}
	\label{fig:handover}
\end{figure}
The sequence diagram in Figure~\ref{fig:handover} illustrates the handover process for runtime verification. The handover process contains the following steps: 
\begin{description}
	\item[\#1] The first step is common to all choreographies. After the sending participant (here: Alice) has selected a potential receiving participant (here: Bob), they mutually identify each other and negotiate the metadata of the handover. Notably, Figure~\ref{fig:handover} simplifies the identification by only showing how Alice identifies Bob, not vice versa.	
	\item[\#2] When a consensus is reached, the sender transfers the symmetrically encrypted process data to the receiver. This way, the time-consuming data transfer is completed before the handover, but the receiver can not yet start working on the following task. On the sender side, the process data is hashed to prove its state during handover.
	\item[\#3] Bitcoin addresses are exchanged through PKI signatures to provide a verifiable confirmation that the address is indeed owned by the respective partner. 
	\item[\#4] A transaction template is created by the sender which holds the negotiated handover terms (i.e., the required metadata to completely describe the state of a process).
	\item[\#5] The sender transmits the transaction template to the receiver together with the symmetric key to unlock the process data. The template is RSA-signed and sent. This way, the receiver has proof that the sender intends to perform the given handover. If the transaction template contains the negotiated handover terms from Step~\#1, the receiver approves the template by creating and returning a Bitcoin-based signature of the template. For this signature, the private key of the receiver's Bitcoin address is used. Since the receiver can now decrypt the process data, the execution of the next process task can be started.
	\item[\#6] The sender validates the receiver's signature. If the signature is correct, the transaction is finalized by adding the Bitcoin-based signature of the sender. At last, the Bitcoin transaction is submitted to the Bitcoin blockchain by the sender. Eventually, the transaction is included in a block, and thus becomes part of the persistent blockchain data store. Since all Bitcoin transactions are broadcast and shared publicly, even before they are included in a block, the receiver can monitor if the sender actually submitted the transaction. If the sender does not do so, the receiver can contact the respective mediator of the choreography (i.e., the process owner). The transaction template, signed by the sender, serves as proof that a process handover was intended by the two partners. 
\end{description}

\begin{figure}[b!]
	\centering
	\includegraphics[width=\columnwidth]{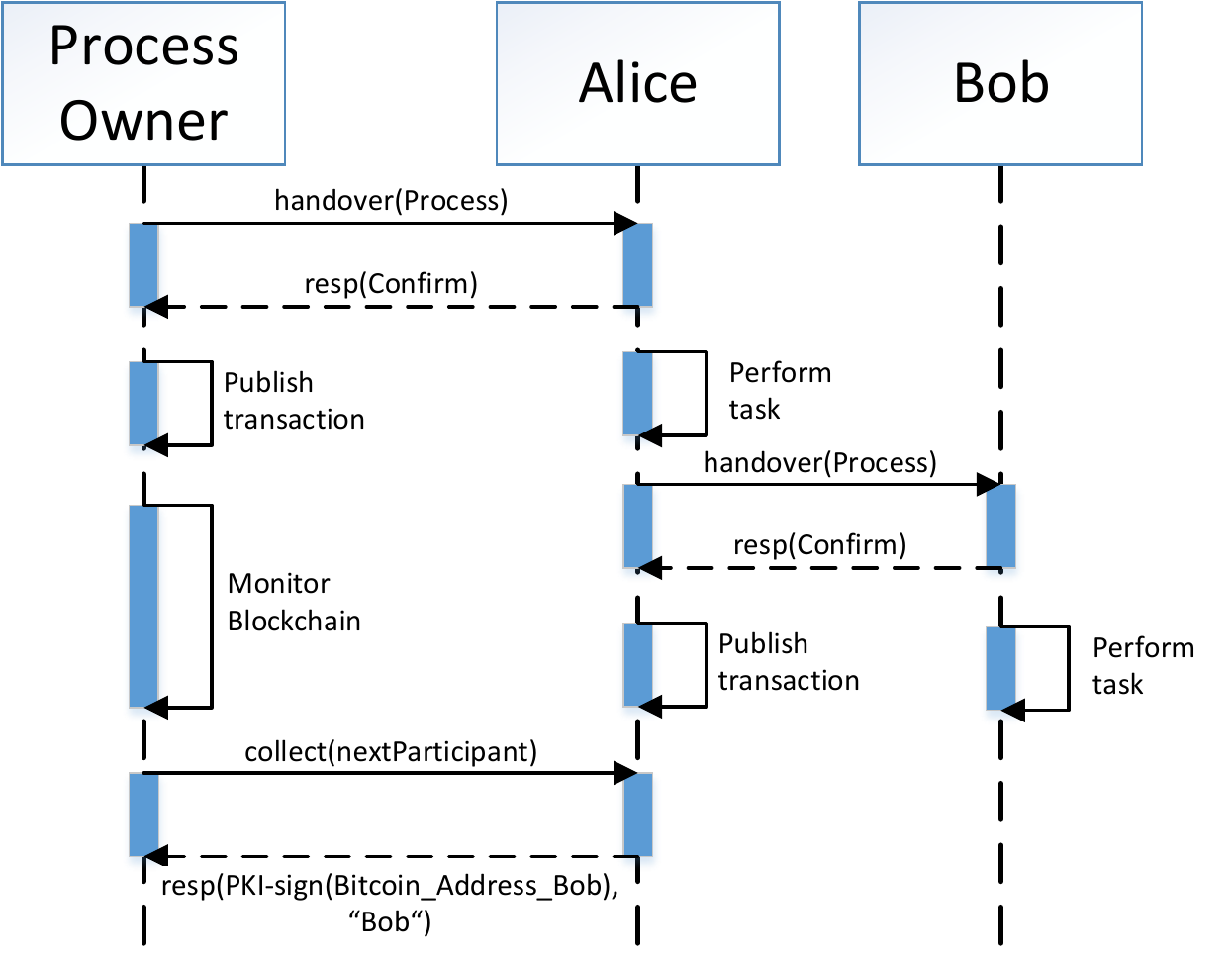}
	\caption{Pull-based Monitoring by Process Owner}
	\label{fig:informationCollectionDetails}
\end{figure}
\begin{figure*}[t]
	\centering
	\subfloat[Process-Handover]{
		\includegraphics[width=0.33\textwidth]{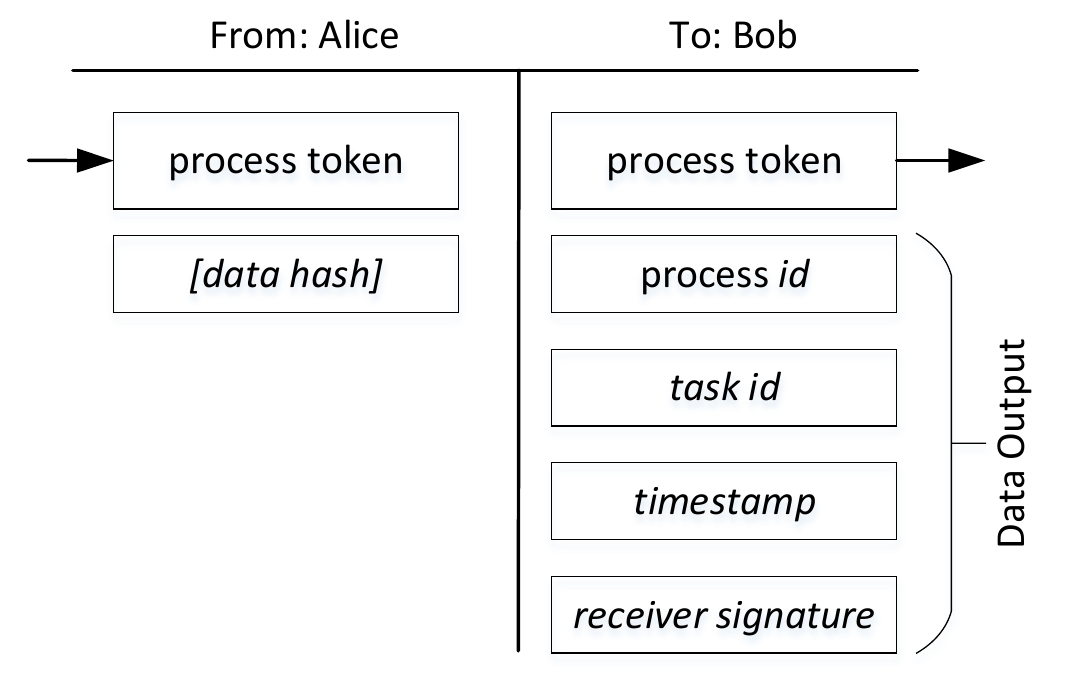}
		\label{fig:processhandover}
	}
	\subfloat[Process-Start]{
		\includegraphics[width=0.33\textwidth]{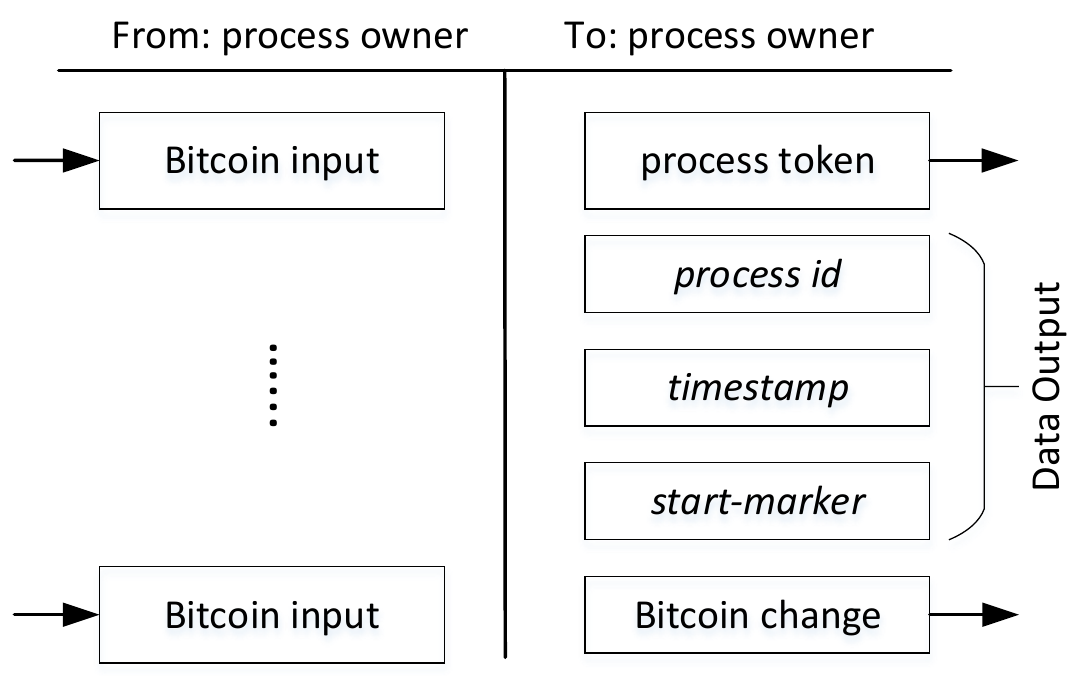}
		\label{fig:processstart}
	}
	\subfloat[Process-End]{
		\includegraphics[width=0.33\textwidth]{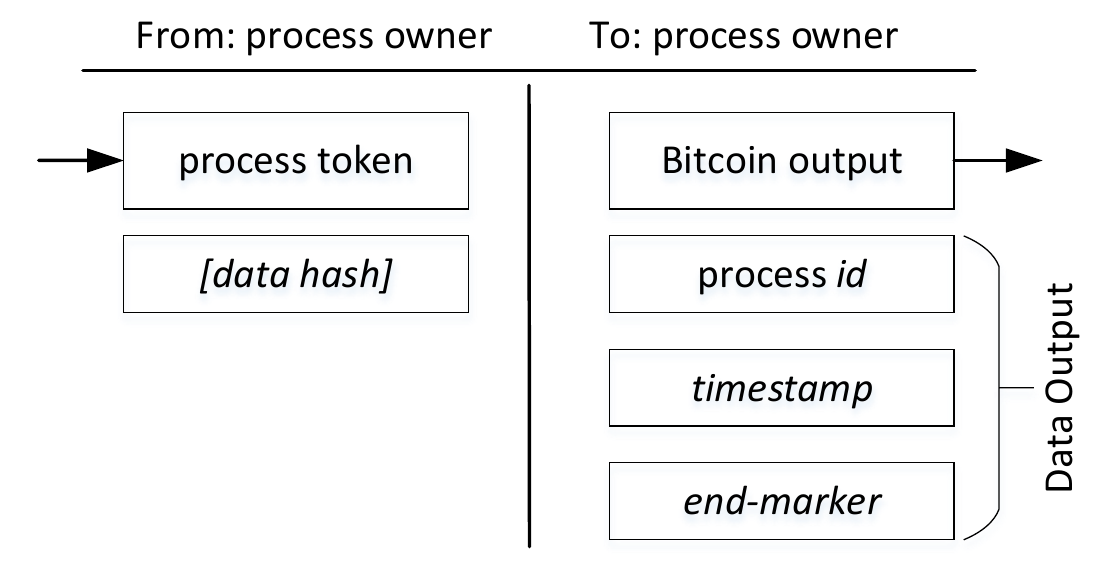}
		\label{fig:processend}
	}\\
	\subfloat[Process-Split]{
		\includegraphics[width=0.33\textwidth]{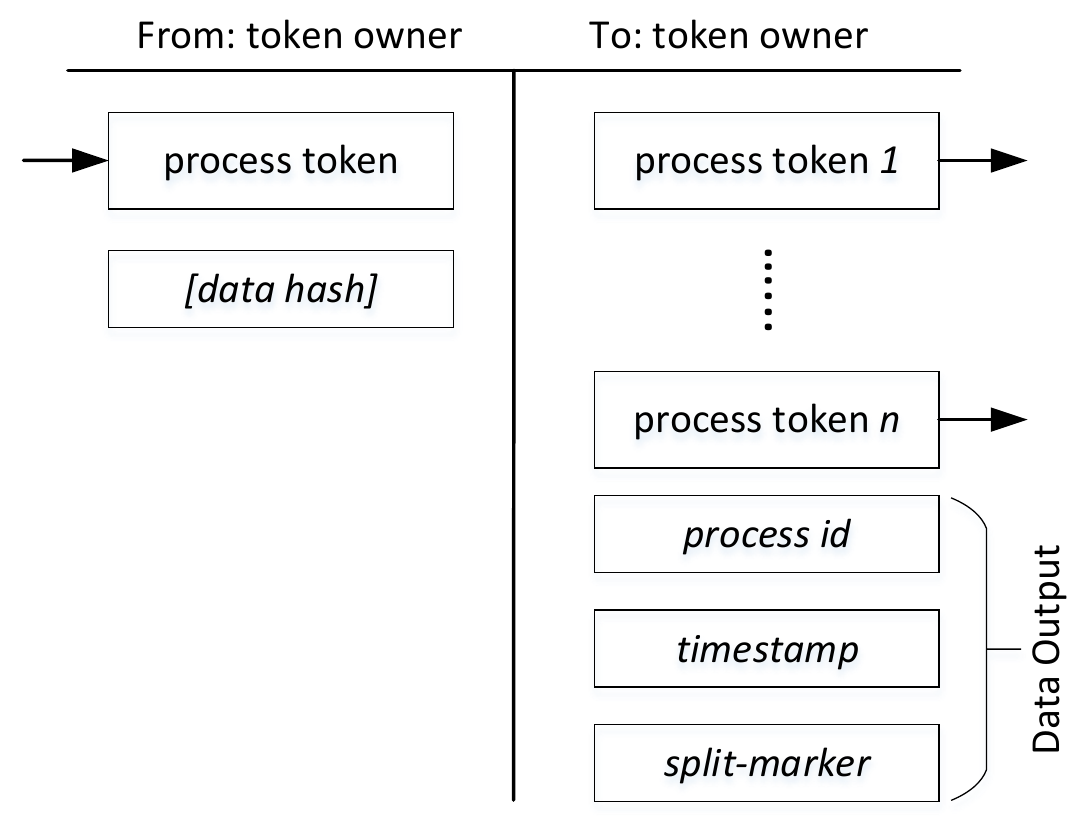}
		\label{fig:processsplit}
	}
	\subfloat[Process-Join]{
		\includegraphics[width=0.33\textwidth]{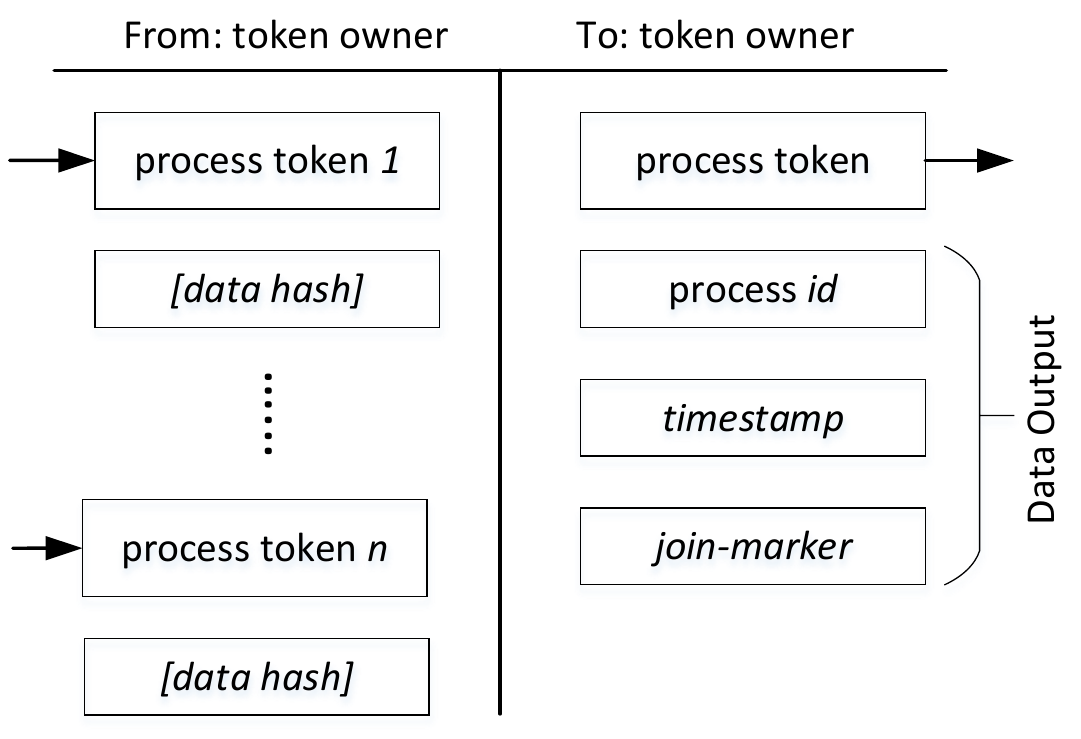}
		\label{fig:processjoin}
	}
	\caption{General Structures of Enriched Transactions}
	\label{fig:transactions}
\end{figure*}

By observing the Bitcoin blockchain, the process owner can monitor the progress of each process instance. Optionally, it is possible for the owner to immediately collect more detailed information about the latest progress of a process instance. For example, initially the process owner hands over the process instance to the first choreography participant, Alice. After completing the designated task, Alice performs a handover to the next participant, Bob. Right after the new handover documentation element is published, the process owner observes the changes and can contact Alice to request Bob's identity information. This information collection is pull-based. This process is further illustrated in Figure~\ref{fig:informationCollectionDetails}. The general structure of a process-handover documentation element is illustrated in Figure \ref{fig:processhandover}. 

Transactions in Bitcoin typically require the payment of a fee, which is specified as the difference between the Bitcoin input amount and the output amount in a transaction. 
In our approach, the process owner supplies sufficient Bitcoins with the process token to cover all transaction fees for the whole process execution.
This enables anonymity of the other participants with respect to the information stored in the blockchain, since they are not required to supply any Bitcoins to pay for transaction fees -- if they were, it would be possible to ``follow the money'' in an attempt to de-anonymize participants. While the size of the fee can be selected autonomously by the participants, it is generally assumed that participants choose the lowest possible fee that still ensures a timely inclusion of the transaction into a new block.

\subsubsection{Further Documentation Elements}
\label{subsub:furthertransactions}
To completely capture the execution of a process instance that includes activities, exclusive-or (XOR) path decisions and parallel execution paths (AND) \cite{vdAalst03a}, additional types of documentation elements (i.e. enriched transactions) are needed. The controlled handover between process partners, as described above, documents the execution of activities. Each handover can mark the end of a previous activity and the start of a new one. XOR path decisions do not require a dedicated documentation marker since they always resolve into one single execution path. By analyzing the sequence of activities, a participant can determine how the XOR path decision was resolved, e.g., by using real-time process conformance checking~\cite{Weber:2015:CCaaS}. Besides this, the following documentation elements are required: (i) start of a process, (ii) end of a process, (iii) split into parallel execution, and (iv) join from parallel execution.

The documentation element to \emph{start a process} is submitted by the process owner. It takes an arbitrary number of Bitcoin inputs and outputs the token to be used for the process instance. It further documents the process id, a timestamp and a specific start-of-process marker. The timestamp of this transaction defines the starting time of its given process instance. At last, a change output can return any surplus Bitcoins. Thus, the transaction corresponding to this documentation element prepares the process instance token. The token itself still remains under the control of the process owner, i.e., the output which holds the token still belongs to the process owner. The general structure of a process-start transaction is illustrated in Figure~\ref{fig:processstart}.

We assume that a process always starts and ends with a dedicated event published by the process owner. Since the process owner usually expects some kind of result to be returned by the participants of the choreography, the final process handover should point back to the process owner. Therefore, the enriched transaction to \emph{end a process}, as shown in Figure \ref{fig:processend}, is also submitted by the process owner. As input, it expects the token of the process instance and optionally a data hash for documentation purposes. The token is no longer required for the runtime verification framework and the output can be used for arbitrary purposes. The transaction furthermore documents the process id, a timestamp and a specific end-of-process marker. The timestamp of this transaction defines the ending time of the given process instance. Input as well as output remain under the control of the process owner. The Bitcoins received through the process token can then be used for new process instances.

To enable parallelism in processes, an enriched transaction to \emph{split a process} is required. This transaction is not meant to be used to transfer process tokens between different participants. Instead, only one participant, who decides to split a process, creates the transaction. The single token input and at least two token outputs are under the control of the current process token holder. The number of outputs defines the number of paths the process execution is split into. For each of the paths, the process token owner can then add individual process-handover transactions to other participants. In the split transaction, the process id, a timestamp, and a specific split-of-process marker are documented. The timestamp of this transaction defines when the given process instance was split into parallel paths. If a data hash was contained from the previous task execution, this data is documented along with the tokens. The general structure of a process-split transaction is illustrated in Figure~\ref{fig:processsplit}.

We assume that every split leads to exactly one corresponding join. To join parallel execution paths, an enriched transaction to \emph{join a process} is required. Similar to the start, end, and split transactions, the process-join transaction is not used to transfer the ownership of a process token. When a choreography participant accepts a process task which requires inputs from multiple execution paths (i.e., it requires a process join), the execution cannot proceed until control over all other execution paths also has been handed over to this specific participant. When executing different process paths in parallel, one path will always be the first one to finish. The process token of this subpath is then handed over to a participant which agrees to proceed the execution. All other participants must then also handover their execution tokens to this specific participant. Finally, this participant merges all execution tokens into a single token, with a dedicated process-join transaction. This transaction expects at least two token inputs with optional data hashes and provides a single token output. As usual, also the process id, a timestamp and a specific join-of-process marker are added. The timestamp of this transaction defines when the given process instance was joined from parallel paths. The general structure of a process-join transaction is illustrated in Figure~\ref{fig:processjoin}.

\subsection{Realizing the Solution with Bitcoin Transactions}
\label{sub:implementation}
As outlined above, we use the Bitcoin blockchain to implement the conceptual solution presented in Section~\ref{sub:approach}. To describe our implementation, first, the different types of Bitcoin transactions are described (Section~\ref{subsub:bitcoinTransactionTypes}). Next, Section~\ref{subsub:handoverTransaction} defines how these transactions are adapted to serve as process verification transactions. 

\subsubsection{Bitcoin Transaction Types}
\label{subsub:bitcoinTransactionTypes}

Only two elements in a Bitcoin transaction do not have a defined length and can be filled with arbitrary contents: The script locking an output, historically called \emph{scriptPubKey}, and the unlocking script provided by an input, historically called \emph{scriptSig}. In this paper, these two scripts will be referred to as \emph{locking script} and \emph{unlocking script}~\cite{antonopoulos14}. 

For these scripts, Bitcoin defines a custom scripting language. While the language is on purpose not Turing-complete, it still enables extensive variants of programs.  
Only if all locking and unlocking scripts of a transaction match one out of five different predefined structures, the received transaction is regarded as ``standard''. This results in five different types of Bitcoin transactions which are solely distinguished by the structure of their locking and unlocking scripts.

Currently, the Bitcoin software discards all incoming transactions that are sent across the Bitcoin network which are not regarded as standard. This means that all miner nodes which run an instance of the Bitcoin software will not accept non-standard transactions. Therefore, to include a non-standard transaction into the Bitcoin blockchain, one must first find a miner that accepts non-standard transactions. In addition, this specific miner then must win the race of creating a new block, which can take a very long time. Although it is not explicitly prohibited to create and publish non-standard transactions, it has become de facto very difficult to integrate them into the blockchain. Hence, the usage of non-standard transactions is not an option for our solution. 

Only two of the five standard transaction types provide the possibility to insert arbitrary data and still be considered as standard \cite{antonopoulos14}. The other three transaction types are called \emph{Pay-to-Public-Key-Hash} (P2PKH), \emph{Pay-to-Public-Key} and \emph{Multi-Signature}. These three transaction types define specific script structures and therefore only serve one specific use case, namely payment. The two standard transaction types which still allow a certain degree of freedom are \emph{data output} and \emph{Pay-to-Script-Hash} (P2SH), which are therefore used in our work. \emph{data output} is the only standard type without a direct purpose for payment. It is designed to serve as simple and limited data field. 
To directly place data in a transaction, a specialized output with zero Bitcoins must be created. Since the only purpose of this output is to contain data, it should never be referenced by an input or carry value. 

To ensure that this output is never successfully consumed by a new input, its locking script must never evaluate to \textit{true}. This is achieved by simply placing the \emph{OP\_RETURN} operator at the top of the script. When this operator is executed, it immediately stops the execution. At that point, the value \emph{true} is not on top of the stack, therefore the validation fails. Miners therefore can safely archive this kind of unspent output and do not have to keep it in memory. As of version 0.11 of the Bitcoin software, it is allowed to store up to 80 bytes of information in such data output and only one data output is allowed per transaction. 

The P2SH type is a powerful transaction type to conduct payments. Amongst others, it allows the controlled usage of non-standard scripts. In contrast to the strictly defined P2PKH transaction type structure, the P2SH transaction type enables the usage of various payment scripts. To allow miners to save memory, the locking script of P2SH transactions remains short and strictly defined. It is the unlocking script which can contain arbitrary data. P2PKH locking scripts require a signature, a public key, and an entire redeem script followed by the redeem script's own parameters as parameters. In contrast, the locking script of a P2SH transaction is a simple hash value comparison. Upon creating a P2SH output, the transaction publisher must decide which script should be provided for unlocking the output and hash it. This hash is then placed in the locking script. 

To unlock an output with such a defined script, a redeem script which matches the placed hash must be provided. In addition, this provided redeem script itself is evaluated and must resolve to true. This feature is normally used for Multi-Signature scripts which tend to be quite long. When used in the P2SH variant, they save memory for the miners. The P2SH validation is then performed in two stages: First, the provided redeem script is compared against the defined hash. Second, the redeem script itself is evaluated with its parameters.

\subsubsection{Handover Transaction Implementation}
\label{subsub:handoverTransaction}
To store the process information required for a handover between two participants in the Bitcoin blockchain, as defined in Section~\ref{subsub:handoverprocess}, the two transaction types \emph{data output} and \textit{P2SH} are used. 
The elements process id, task id, timestamp and receiver signature are included in a transaction by using a data output element. By using a simple adapted P2SH output, the process data hash can also be included. The only downside of using P2SH elements is that the stored data can only be placed in the redeem script. This redeem script is part of the unlocking script which becomes only visible on the blockchain after the output has been spent, i.e., another transaction consumed the output by placing the redeem script on the blockchain.

\begin{figure}[tb]
	\centering
	\includegraphics[width=\columnwidth]{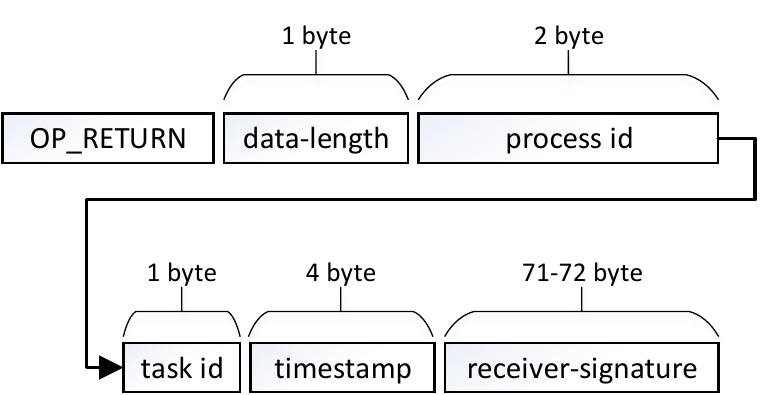}
	\caption{Structure of Process Data in a Bitcoin Data Output}
	\label{fig:opReturnTransactionForprocess}
\end{figure}

The 80 bytes storage of the data output element are divided as depicted in Figure~\ref{fig:opReturnTransactionForprocess}: The first byte is used to store the length of the stored data block, which may vary depending on the process transaction type. Next, two bytes are reserved to store the process instance id. This enables the definition of $65,535$ unique process instances in this kind of runtime verification. To store the task id, one byte is reserved. Therefore, $255$ different tasks can be defined inside a process model. The task id is followed by the timestamp with a length of 4 byte. At last, the Bitcoin signature of the receiver is placed. The length of this signature is not fixed but commonly ranges between $71$-$72$~bytes. This results in a data block with a total length of $79$-$80$~bytes. 

The process data hash does not fit into the data output element. Therefore, a P2SH transaction has to be used to store it. At the same time, such P2SH outputs are used to transfer process instance tokens. Therefore, also the access restriction features of a P2PKH transaction are required. To achieve this, a P2PKH script with an optional data hash is used as redeem script. The data hash itself does not add any specific functionality to the script, so we just place it on the blockchain as a string of characters. Bitcoin provides a script command (\emph{OP\_DROP}), which we use to ensure that the hash is removed from the stack before the actual P2PKH script is executed. The inclusion of the data hash is optional. 

As mentioned before, the process data hash is placed on the blockchain only after the given output has been consumed, i.e., the given token has been passed on. The redeem script hash placed in the P2SH locking script ensures that only a correct process data hash can be placed in the redeem script. Furthermore, it is guaranteed that before a handover transaction can be published, both process participants are in possession of the process data in its current state. The receiver of a process handover is able to verify that the given P2SH output incorporates the correct process data hash. In addition, the sender of a process handover can provide the data hash when the process owner demands it. This way, a process owner can also verify that a data hash has been stored, even if the corresponding token output has not yet been passed on.

The general steps to perform a handover between two participants have already been illustrated in Figure~\ref{fig:handover}. In Step~\#4, a \emph{handover-transaction template} is created by the sender of the handover which includes almost all required process data. This template is only missing two signatures, one from the receiver and one from the sender. Upon receiving the template, the handover receiver is able to validate the correctness of the following critical requirements:
\begin{description}
	\item[INPUT\#1 contains redeem script:] INPUT\#1 references the output of a previous transaction. This output must contain a P2SH locking script. The unlocking script, currently included in the template, is still missing parameters (i.e., the sender signature and public key) but the redeem script itself is already complete. Therefore, the handover receiver is able to validate the correctness of the redeem script, thus also the correctness of the included data hash of the last transaction.
	\item[OUTPUT\#1 can be retrieved:] Though the data of OUTPUT\#1 is abstracted by a P2SH script, the handover receiver knows the defined structure of the script. In addition, all required data to check the correctness of the included script hash is already known. The redeem script is constructed from the hash of the receiver's public key and the hash of the already transferred process data. By recreating the redeem script and comparing it to the hash placed in OUTPUT\#1, the handover receiver ensures that the token is indeed correctly passed on and that the hash of the just received process data is correct.
	\item[OUTPUT\#2 contains the negotiated process data:] Except for the process data hash, all negotiated data is included in OUTPUT\#2, as shown in Figure~\ref{fig:opReturnTransactionForprocess}. Therefore, the correctness of the included data can be directly verified.
	\item[Previous process execution is valid:] The handover receiver also obtains the process model. Since INPUT\#1 is referencing a previous process transaction, the receiver is able to trace the execution history of the process instance. Besides other meta-information about the process, it can be determined if the process execution still conforms with the defined process model.
\end{description}
If the received handover template is successfully verified, the handover receiver signs it. For the signature, the same Bitcoin key-pair is used that has been utilized to receive the token in OUTPUT\#1. The technical structure of a process-handover transaction is illustrated in Figure \ref{fig:processhandover} which follows the general description of the transaction provided in Section~\ref{subsub:handoverprocess}.

It is also possible to incorporate other types of redeem scripts inside the P2SH script. Fault management strategies can be employed through the usage of the Multi-Signature type script instead of the P2PKH type script. While the P2PKH script can only be configured to be unlocked by the new token owner, a Multi-Signature script can be unlocked by multiple different parties. In case a fault occurs during a process, an escalation strategy has to be employed by the process owner. To enable the process owner to intercept a process token in the case of incorrect execution, a Multi-Signature script can be placed as redeem script in the P2SH locking script. The script then grants access for two parties instead of one, namely the next token owner and the process owner. For instance, if the new token owner stops responding and does not perform the negotiated task, the process owner could decide to collect the token instead. However,, fault management is out of scope for the work at hand. Therefore, in the proposed prototype only simple P2PKH scripts are employed as redeem scripts.

\subsection{Runtime Verification Framework and Implementation}
\label{sub:framework}
We developed a Java-based software framework that implements the runtime verification approach as described above. This framework is designed to abstract all implementation details to simplify its integration into any arbitrary choreography-oriented BPMS. The framework is further enhanced by using a \emph{simple payment verification} (SPV) Bitcoin client as slim foundation, a \emph{remote REST API} for data collection, and a \emph{greedy publishing} mode. While these features are not essential for the implementation of the proposed concepts, they increase the usability and testability of the prototype. 

During the handover of a choreography, the receiving participants want to verify the execution path of the instance prior to the handover. Therefore, they need to be able to obtain information about old transactions that have been submitted to the network or are currently pending. Since the implementation proposed by this work relies on a SPV foundation, only block hashes and transactions directly related to the client's wallet are monitored. To obtain information about transactions unrelated to the wallet, a separate data collection framework has to be used. Many different companies, e.g., blockchain.info\footnote{\url{https://blockchain.info/de/api}}, Blockcypher\footnote{\url{https://api.blockcypher.com}}, or Blocktrail\footnote{\url{https://www.blocktrail.com/api}} provide live access to Bitcoin information through remote REST APIs. By calling these REST interfaces, slim Bitcoin clients can obtain information about any confirmed or still pending transaction in the Bitcoin network. This way, clients are able to reconstruct the execution of a process instance without running a full Bitcoin node. For the prototypical implementation in this paper, the REST API of Blockcypher is used.

The required proof of work for the creation of new blocks in Bitcoin is configured to result in a median block creation time of ten minutes. Unfortunately, there is a lot of variation in the time between block creations. The time between the arrival of two blocks roughly follows an exponential distribution. There may be mere seconds between the creation of two blocks or even an hour~\cite{franco14}. If too many transactions are published at the same time or if blocks are created too slowly, published transactions must be buffered by the miners of the Bitcoin network. On average, approximately 1,500 transactions are currently included in a new block.\footnote{\url{https://blockchain.info/charts/n-transactions-per-block?timespan=1year}, accessed: 15.01.2017} If there are more than 1,500 transactions queued to be included in a new block, some transactions might need to wait two or even three block creation intervals until they are confirmed. Also, the fees offered by each transaction affect their confirmation time. Transactions with higher fees are treated with higher priority. This results in an even higher variation for the confirmation time of transactions~\cite{franco14}.

Therefore, the transaction confirmation time is expected to be a major challenge for a runtime verification approach based on the Bitcoin blockchain. The conservative approach to runtime verification would be to wait for the confirmation of each published transaction before appending new transactions to it. For long-running processes with long intervals between handovers, like supply chain processes, this is sufficient. But it constitutes a bottleneck for fast-running processes with short tasks between handovers, like software computations. 

Therefore, we include a \emph{greedy} publishing mode into our framework. This greedy approach allows to append new transactions to a process, even though the latest ones have not been confirmed yet. Nevertheless, since these transactions are broadcast across the Bitcoin network and buffered by the miners, they can be retrieved by process participants. Choreography participants are still able to verify the validity of a transaction for a given process instance. Also, Bitcoin miners accept new transactions which reference unconfirmed transactions as input. It is possible to create whole chains of unconfirmed transactions, enabling the participants of a shared process instance to continue with the process execution even though not all process transactions have been included into the blockchain yet.

By using this approach, the execution of a fast-pacing process is not delayed. Furthermore, multiple chained transactions can be confirmed at once in a single block. Through this method, less blocks have to be created to confirm all published transactions of a process instance. In addition, each participant is in possession of the complete process execution chain and can prove that handovers of the process instance took place that were confirmed by both handover partners. The risk of this approach is that a whole chain of pending transaction may be dropped if something goes wrong. 

If one intermediate transaction of the pending chain is somehow lost, the whole remaining chain is also dropped since it became invalid. Each published transaction is \emph{flooded} through the Bitcoin network and stored in the buffers of various miners. 

Even during a conflict in the blockchain, where it may occur that single blocks are dropped, the transactions should still remain in the pending buffer. The highest threat for this greedy approach is malicious behavior of one of the participants. Each participant is theoretically able to publish an alternative version of an intermediate transaction in the pending chain. In this case, two alternative transactions become available for a single output. It is undefined which transaction will be included in the blockchain. If the alternative transaction of the malicious participant is chosen, the remaining pending transaction chain is dropped. On the contrary, it is very unlikely that the alternative transaction, published by the malicious participant, is a valid handover transaction. The malicious participant still requires another participant to confirm the handover. At the same time, this participant can easily verify that another process transaction is already pending for the given token output. At last, the alternative transaction also documents which participant caused the disruption, resulting in penalties and loss of reputation. The definition and implementation of such fault management and fallback mechanisms remains part of our future work.

Greedy publishing aims to enable fast-paced processes in the slow running environment of the Bitcoin blockchain. To enable this feature in conjunction with an SPV client also requires the usage of a remote REST API to fetch additional information. If a full node would be used, the feature of greedy publishing would still be possible but the usage of the remote REST API could be reduced. Since the full node receives and buffers most pending transactions itself, the REST API would only be required for exceptions.

\newpage
\section{Evaluation}
\label{sec:evaluation}
To evaluate the applicability of the presented solution for choreography runtime verification, we implemented our approach as a research prototype. The prototype\footnote{The software can be downloaded at \url{https://github.com/ChristophPrybilaTUVienna/ChoreographyRuntimeVerificationByBlockchain}.} was implemented using Java JDK 1.8, Apache Maven 3.3.9 for dependency management, Spring Beans 4.2.6 for dependency injections, Apache HttpClient 4.5.2 and Google Gson 2.7 for the REST API, BitcoinJ 0.14.2 as the Bitcoin API, and JUnit 4.12 for testing purposes. 

We evaluate our work in terms of a qualitative comparison to existing solutions (Section~\ref{sub:qualitative}) and conduct a performance evaluation (Section~\ref{sub:performance}). 

\subsection{Qualitative Evaluation}
\label{sub:qualitative}

Choreographies operate as decentralized systems. The more heterogeneous, geographically decentralized, and organizationally independent such systems become, the more dynamic and diverse they are. Therefore, in choreography-oriented BPM, many unique situations have to be foreseen by a verification approach~\cite{leite12,montagut08,fdhila12}. For instance, process participants might become unavailable, tasks might be processed incorrectly, or participants might compete with each other. In order not to limit according fault management strategies, a runtime verification system must remain as flexible as possible. In this paper, the term \emph{flexibility} is therefore used to describe the capability of a runtime verification system to deal with the dynamic nature of decentralized choreographies and the process participants.

\begin{table*}[th]
	\footnotesize
	\centering
	\caption{Criteria Application to State of the Art Approaches}
	\label{tab:comparison}
	\begin{tabularx}{\textwidth}{lccccccX}
		\toprule
		& Participant & Participant & Data & Internal Structure & Discovery of & Sequence & Verification\\
		& Selection & Identities Sharing & Sharing & Sharing & Incorrect Behavior & Enforcement & Trust Basis\\
		\midrule
		Bengtsson et al., 2005~\cite{bengtsson05} & On-Demand & All & All & Not required & Semi-supported & Not Supported & Signature-enhanced Token\\
		Montagut et al., 2008~\cite{montagut08} & Predefined & Minimum & Minimum & Not required & Semi-supported & Supported & Cryptographic Onion\\
		Von Riegen et al., 2009~\cite{vonriegen09} & On-Demand & Minimum & Minimum & Unknown & Supported & Supported & Enterprise Service Bus\\
		Baouab et al., 2011~\cite{baouab11} & Predefined & Minimum & Minimum & Unknown & Semi-supported& Supported & Message\\
		& & & & & & & Interception\\
		Lim et al., 2012~\cite{lim12} & On-Demand & All & All & Not required & Semi-supported& Not Supported & Hierarchical Signatures\\
		Hwang et al., 2013~\cite{hwang13} & On-Demand & Minimum & Minimum & Not required & Supported& Supported & Centralized Cloud\\
		& & & & & & & Storage\\
		Weber et al,. 2016~\cite{weber16} & Predefined & Minimum & Mixed & Not required & Supported & Supported & Blockchain \\
		\midrule
		This Work, 2017& On-Demand & Minimum & All & Not required & Supported & Not Supported & Blockchain \\
		\bottomrule
	\end{tabularx}
	\normalsize
\end{table*}

Following approaches for qualitative comparison as presented, e.g., in~\cite{hofer11,annette15}, we have extracted the following comparison criteria for choreography runtime verification. Table~\ref{tab:comparison} shows how approaches from the state of the art (introduced in Section~\ref{sec:related}) and our own solution fulfill the discussed criteria. In addition, the technical approaches to realize a verification trust basis are named.
\begin{description}
	\item[Participant Selection.] A major influence on the stability of a decentralized process instance is the selection of the participants~\cite{montagut08,fdhila12}. Predefining process participants \emph{before} runtime greatly reduces the organizational efforts during process execution. At the same time, a process instance becomes less robust, since participants cannot be replaced during runtime. In a predefined setting, a process instance may halt in this case. An alternative is to select the required participants ad hoc during process execution. Especially for long-running processes where participants may have to wait a long time before they are involved, this flexible approach can increase the overall robustness.
	\item[Information Sharing.] If choreography-oriented process execution takes place as cooperation between independent participants, information sharing becomes an issue~\cite{baouab11,vonriegen09}. In particular, participants of B2B choreographies might even be competitors. Therefore, it becomes a requirement to share as little information as possible with other participants. With regard to the work at hand, it is important to evaluate which internal information is absolutely necessary. The less information is shared, the more acceptable the system becomes for independent participants.
	
	Some state of the art runtime verification approaches require that the \textit{identities of all participants} are shared between all process participants, while other approaches explicitly foresee that only minimum information about process participants is shared. Analogue considerations apply with regard to general \textit{data sharing}, i.e., that not all process-related data needs to be shared with all process participants, and with regard to (company-)\emph{internal structure sharing}.
We discuss these concerns in the following.
	
	In a choreography, the process owner has to know the identities of all participants. Since a process instance is passed along the participants, in general each participant knows at least two other participants, i.e., the previous and the subsequent process participants (although they may be the same in some cases). We refer to this setup as ``Minimum'' (see Table~\ref{tab:comparison}) with regard to \emph{participant identities sharing}. Some verification approaches require the complete sharing of identities between the participants, which we define as sharing ``All'' participant identities.
	
	With regard to \emph{data sharing}, the process owner will in most cases send some data along with the process instance. Each process task may require and alter this data. Since each task may be executed by a different process participant, a given participant may only know the state of the process data before and after the task has been executed. This setup can be defined as ``Minimum'' for process data sharing. Some verification approaches enable all participants to view the content, or trace the complete evolution of the shared data, which can be defined as sharing ``All'' data whereas ``Mixed'' is in between these two extremes.
	
	With regard to \emph{internal structure sharing}, approaches are graded according to their invasiveness. A purely token-based approach is considered very lightweight, since no dedicated infrastructure needs to be installed at the participants. Message-based approaches require dedicated resources (e.g., a service bus, filtering components). Some of these resources are operated/provided by third parties. If they are integrated into the BPMS of a process participant, this third party requires internal structure sharing. 
	\item[Correctness Verification.] A runtime verification framework should support the discovery of incorrect process execution. It must not be possible for a participant to conduct \textit{incorrect behavior} during the execution of a choreography-oriented process instance without being noticed by the process owner or other process participants. We refer to an approach from the related work as ``semi-supported'' when that incorrect behavior can be discovered in ideal scenarios, but can become impossible in situations like the loss of a process token. In addition to this feature, some runtime verification frameworks are able to apply even stricter constraints: the \textit{enforcement of execution sequences} for process instances means that illegitimately ordered tasks are rejected by the system.
\end{description}
When comparing our own approach with the related work (see Table~\ref{tab:comparison}), two aspects require further explanation. First, our approach requires a higher degree of data sharing than existing approaches. In fact, restricting data sharing has not been in the focus of our work, therefore all process-related data is shared with the other process participants. Second, since process models are not incorporated in the logic of a process transaction, the sequence of performed tasks can not be enforced. However, it is possible for the process owner to realize \emph{ex post} if the correct process path has been taken by all participants.

In summary, the perfect approach for runtime verification in choreography-oriented BPM has not been found yet. Many approaches utilize tokens that are passed along the participants during process execution to document and/or control the progress of a choreography \cite{montagut08,bengtsson05,lim12,hwang13}. In token-only approaches, participants have to be predefined to subsequently protect the privacy of identity and data. Furthermore, the execution sequence can only be enforced in this way. All token-only approaches suffer from limited discovery of incorrect behavior. Since the token is the only element of proof that work was done, the communication of this token is critical. A token might get intercepted or lost. Finally, a process owner is only able to monitor the progress of a choreography if the intermediary token is returned periodically.

Other approaches aim to control the messages that are passed between the participants \cite{baouab11,vonriegen09,hwang13}. To enable this, different communication and control facilities have to be established. As depicted in Table~\ref{tab:comparison}, these proposals provide a number of benefits. On the downside, the additionally required facilities become the new shared trust basis. If the facilities are owned by a particular entity, then all participants need to trust this entity.

Recent approaches like the one presented by Weber et al.~\cite{weber16} or the work at hand aim to overcome the issue of the shared trust basis by incorporating a blockchain. On this basis, novel solutions are able to provide complete discovery of incorrect behavior and to fully protect the internal technical structure of the participants. At the same time, the respective blockchain serves as secure, independent and decentralized basis of trust.

Weber et al. \cite{weber16} require the participants to be selected in advance. In turn, the proposed solution is able to partially protect the confidentiality of process data and to enforce the execution sequence of the processes. Our approach exhibits greater flexibility by enabling an on-demand participant selection. However, our approach is not able to provide confidentiality for the shared data and the correct execution sequence can not be enforced. In addition, the approaches differ in terms of the used blockchain technology, as discussed in Section~\ref{sec:related}.

\subsection{Performance Evaluation}
\label{sub:performance}
To assess the impact of the runtime verification feature on the performance of choreographies, we have conducted various experiments with the implemented solution. This way, an estimation about the solution's impact on process instances can be made. In the following subsections, we describe our evaluation setup (Section~\ref{subsub:setup}), the evaluation results (Section~\ref{subsub:results}), and discuss the results (Section~\ref{subsub:discussion}).
\subsubsection{Evaluation Setup}
\label{subsub:setup}
\begin{table}[th]
	\footnotesize
	\centering
	\caption{Characteristics of Evaluation Process Models}
	\label{tab:models}
	\begin{tabularx}{0.66\columnwidth}{cYYY}
		\toprule
		Process No. & $|$Steps$|$ & $|XOR|$ & $|AND|$ \\
		\midrule
		\#1 & 3 & 0 & 0 \\
		\#2 & 4 & 1 & 0 \\
		\#3 & 4 & 0 & 1 \\
		\#4 & 5 & 1 & 1 \\
		\bottomrule
	\end{tabularx}
	\normalsize
\end{table}
We evaluate the performance of our approach with regard to three distinct factors: (i) The cost occurring due to the usage of the Bitcoin blockchain as a decentralized, trustworthy ledger which documents the process transactions, (ii) the runtime performance overhead of our verification approach (compared to evaluation runs without verification), and (iii) the ability of the framework to detect incorrect executions.

To measure if the structure of a choreography influences the runtime behavior of our verification approach, four different process models have been applied in our evaluation. In general, every business process model which contains at least two tasks is a suitable candidate for a runtime verification system. To maximize the verification effort for our prototype, every task is carried out by a distinct participant. Table~\ref{tab:models} shows the basic characteristics of the evaluated process models, i.e., how many steps they contain and whether they include AND-blocks, XOR-blocks, or a combination of both. AND-blocks and XOR-blocks cover both splits and merges in the process models.

To start, control, and end a single evaluation run, a small Java application operates as the main controlling entity. In order to deliver deterministic results, the application expects a number of input parameters to configure the evaluation runs, namely: (i) a \textit{process model ID}, (ii) a definition of the different \textit{variants}, e.g., if an XOR gateway is part of a model, 
(iii) a \emph{verification flag} which determines if runtime verification should be carried out or not, and
(iv) a \emph{greedy flag} which determines if the greedy mode as discussed in Section~\ref{sub:framework} is applied. 
Using these parameters, we are able to set the path through a process model, the participants who are responsible for carrying out specific tasks, and what data is produced by certain tasks.

\subsubsection{Evaluation Results}
\label{subsub:results}
In total, 122 process instances were executed, leading to the publication of 450 transactions enriched with process metadata to the Bitcoin blockchain. The total execution time of all process instances amounts to 50.551 hours, which includes waiting periods for transaction confirmations. Each published transaction must reach at least a confirmation depth of 1 before an execution is considered finished. 

Approximately $0.085417$~Bitcoins were spent on fees for the transactions of the experiment. On 31 August 2016, the currency exchange rate of Bitcoin (BTC) to Euro was 1 BTC =~\euro~$512.8969$.\footnote{\url{http://api.coindesk.com/v1/bpi/historical/close.json?currency=EUR&start=2016-08-31&end=2016-08-31}} The cost to fuel the transactions in fiat currency therefore amounts to approximately \euro~$43.81$. Considering that $450$ transactions were published during the simulation, the average fee of a Bitcoin transaction enriched with process metadata results to $0.000189816$~BTC or \euro~$0.09735581$.

To create a baseline for the impact of our proposed runtime verification framework, the business process models described in Table~\ref{tab:models} were executed as choreographies without the blockchain-based runtime verification. Table~\ref{tab:verificationLessRuns} presents the resulting mean execution times of the different evaluation runs.
Without the runtime verification framework, the duration of the evaluation runs is very consistent. In comparison to the mean duration, the standard deviation ($\sigma$) is very small. These baseline values are compared to evaluation runs with the runtime verification framework included. 

We then executed evaluation runs with and without the proposed greedy mode enabled. We expected that these evaluation runs exhibit a higher standard deviation due to their dependency on the Bitcoin blockchain. 

\begin{table}[b]
	\centering
	\caption{Verification-less Process Runs (Baseline, without Blockchain)}
	\footnotesize
	\begin{tabularx}{\columnwidth}{ccYYY}
		\toprule
		Process & Tasks & Evaluation & Mean & Standard \\
		No. & Covered & Runs & Duration [s] & Deviation ($\sigma$)\\
		\midrule
		\#1 & 3 & 3 & 15.544 & 0.074\\
		\#1 & 2 & 3 & 10.468 & 0.012\\	
		\#2 & 3 & 3 & 15.510 & 0.048\\
		\#2 & 3 & 3 & 15.553 & 0.130\\	
		\#2 & 1 & 3 & 5.016 & 0.001\\
		\#3 & 4 & 3 & 18.016 & 0\\
		\#3 & 4 & 3 & 18.016 & 0\\
		\#3 & 3 & 3 & 13.021 & 0.007\\
		\#4 & 4 & 3 & 15.592 & 0.049\\
		\#4 & 4 & 3 & 15.548 & 0.015\\
		\#4 & 4 & 3 & 16.570 & 1.400\\
		\#4 & 4 & 3 & 15.539 & 0.023\\
		\bottomrule
	\end{tabularx}
	\normalsize
	\label{tab:verificationLessRuns} 
\end{table}

\begin{table}[th]
	\footnotesize
	\caption{Verification Framework Process Evaluation Runs (using Blockchain)}
	\subfloat[In non-Greedy Mode]{
		\begin{tabularx}{\columnwidth}{ccccYY}
			\toprule
			Process & Tasks & Evaluation & Total \# & Mean & Standard \\
			No. & Covered & Runs & of Tasks & Duration [s] & Deviation ($\sigma$)\\
			\midrule
			\#1 & 6 & 2 & 12 & 5,501 & 936\\
			\#2 & 6 & 4 & 24 & 3,489 & 1,110\\
			\#3 & 10 & 4 & 40 & 7,835 & 2,309\\
			\#4 & 10 & 4 & 40 & 9,974 & 6,136\\
			\#4 & 9-10 & 4 & 37 & 7,538 & 3,444\\
			\#4 & 9 & 4 & 36 & 9,592 & 3,107\\
			\midrule
			\multicolumn{3}{r}{Total} & 189 & & \\
			\bottomrule
			\label{tab:verificationFrameworkRunsNonGreedy} 
		\end{tabularx}	
	}\\
	\subfloat[In Greedy Mode]{
		\begin{tabularx}{\columnwidth}{ccccYY}
			\toprule
			Process & Tasks & Evaluation & Total \# & Mean & Standard \\
			No. & Covered & Runs & of Tasks & Duration [s] & Deviation ($\sigma$)\\
			\midrule
			\#1 & 6 & 4 & 24 & 707& 430\\
			\#1 & 3 & 6 & 18 & 650& 945\\
			\#2 & 5-6 & 4 & 21 & 711 & 306\\
			\#2 & 4 & 5 & 20 & 4,050 & 7,251\\
			\#3 & 9-10 & 4 & 39 & 542 & 221\\
			\#4 & 10 & 3 & 30 & 2,830 & 2,040\\
			\#4 & 11 & 5 & 55 & 1,155 & 646\\
			\midrule
			\multicolumn{3}{r}{Total} & 207 & & \\
			\bottomrule
			\label{tab:verificationFrameworkRunsGreedy} 
		\end{tabularx}
	}
\end{table}

In order to demonstrate that the proposed runtime verification framework is capable of detecting errors, a number of processes include corrupted handovers. 
Hence, not all evaluation runs are suited for the overall time impact comparison of the runtime verification prototype. Therefore, not all $32$ evaluation run configurations we have applied are listed in Tables~\ref{tab:verificationLessRuns}, \ref{tab:verificationFrameworkRunsNonGreedy} and \ref{tab:verificationFrameworkRunsGreedy}, and thus the numbers of transactions in Tables~\ref{tab:verificationFrameworkRunsNonGreedy} and \ref{tab:verificationFrameworkRunsGreedy} do not amount to the total of $450$ submitted transactions.

The following paragraphs discuss the results of the runtime verification framework without the greedy mode. Afterwards, the impact of the runtime verification framework if the greedy mode is activated is also analyzed.

\paragraph{Non-Greedy Mode Results}

\begin{figure}[t]
	\centering
	\includegraphics[width=0.9\columnwidth]{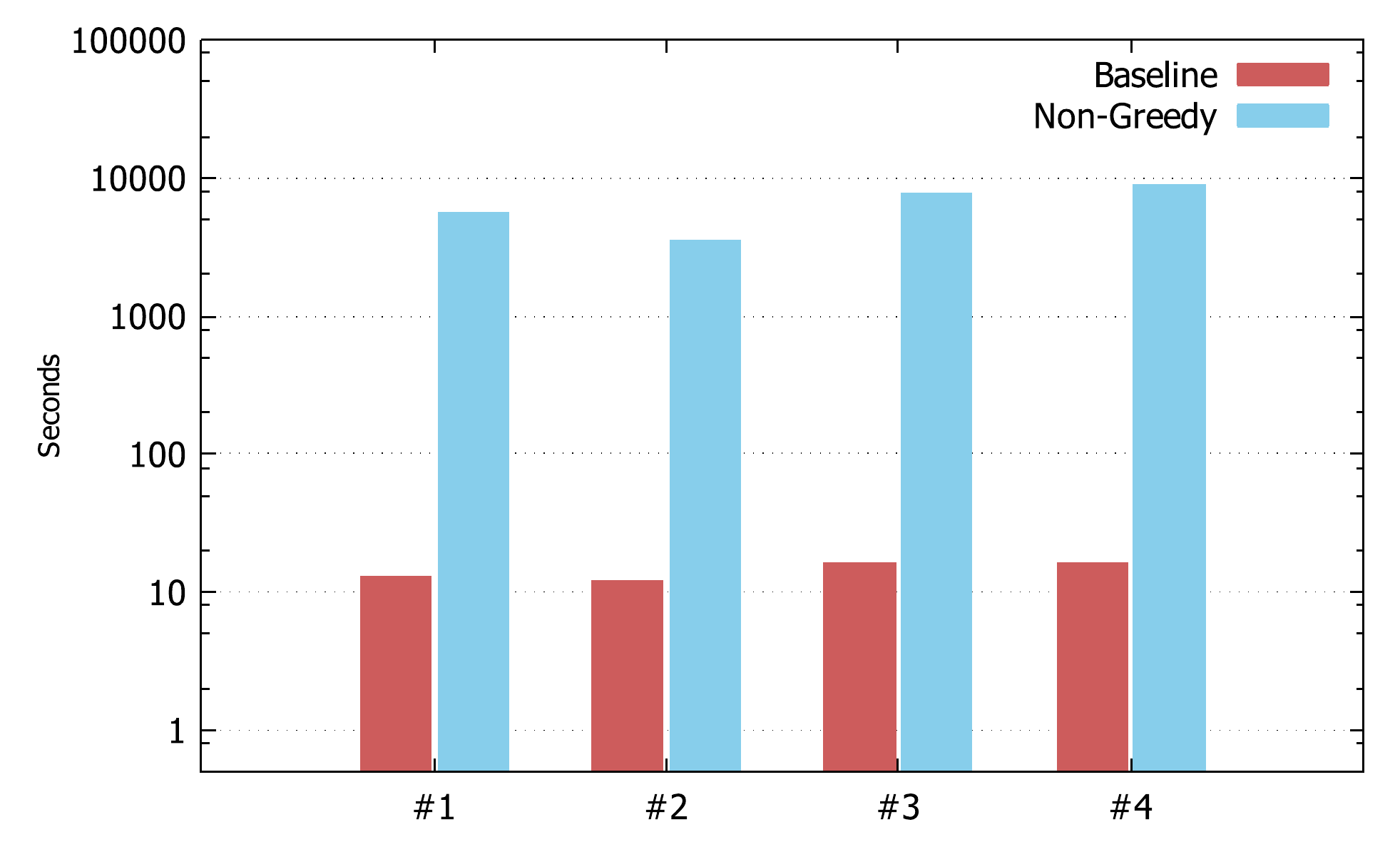}
	\caption{Bar Chart Comparing Durations of the Baseline and the non-Greedy Evaluation Runs}
	\label{fig:DurationComparison_Sync_BarPlot}
\end{figure}
Table~\ref{tab:verificationFrameworkRunsNonGreedy} shows the mean execution times of the runtime verification-based evaluation runs without greedy mode. Figure~\ref{fig:DurationComparison_Sync_BarPlot} shows a comparison of the durations of the baseline without verification and the non-greedy approach. The large overhead caused by the runtime verification can be partially traced back to REST API requests and logical tasks carried out by the framework, but the biggest part is the duration of the transaction confirmation. In further tests, we have measured that the latter is responsible for 99.42\% of the mean duration of non-greedy runtime verification steps, while the framework logic takes only 0.05\% and the REST API requests take 0.53\%. These results show that waiting for a transaction to confirm takes up the most time if using the runtime verification framework in non-greedy mode. Still, knowing the mean duration for a process task does not enable estimations of the expected increase in overall execution duration. This becomes obvious when the mean evaluation run durations from Tables~\ref{tab:verificationFrameworkRunsNonGreedy} and \ref{tab:verificationFrameworkRunsGreedy} are compared with respect to their respective standard deviations $\sigma$. The execution durations of both modes have a very high standard deviation in comparison to their means. 
Therefore, even if the number of required recording steps of a process instance is known in advance, the exhibited execution duration varies significantly. 

\begin{figure}[b]
	\centering
	\subfloat[Non-Greedy Transactions]{
		\includegraphics[height=5.25cm]{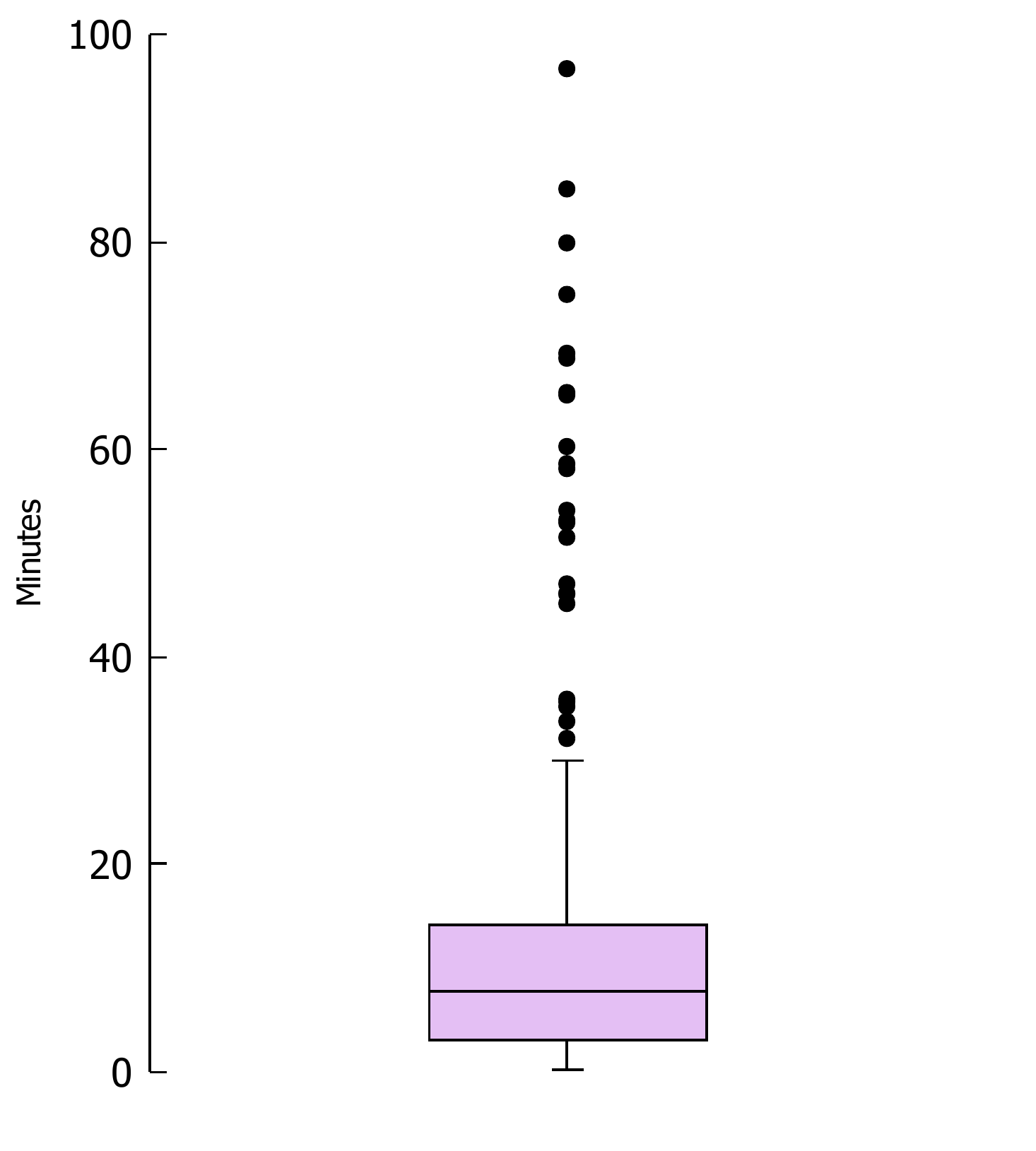}
		\label{fig:TxConfirmationDurations_Sync_Boxplot}}
	\subfloat[Greedy Transactions]{\includegraphics[height=5.25cm]{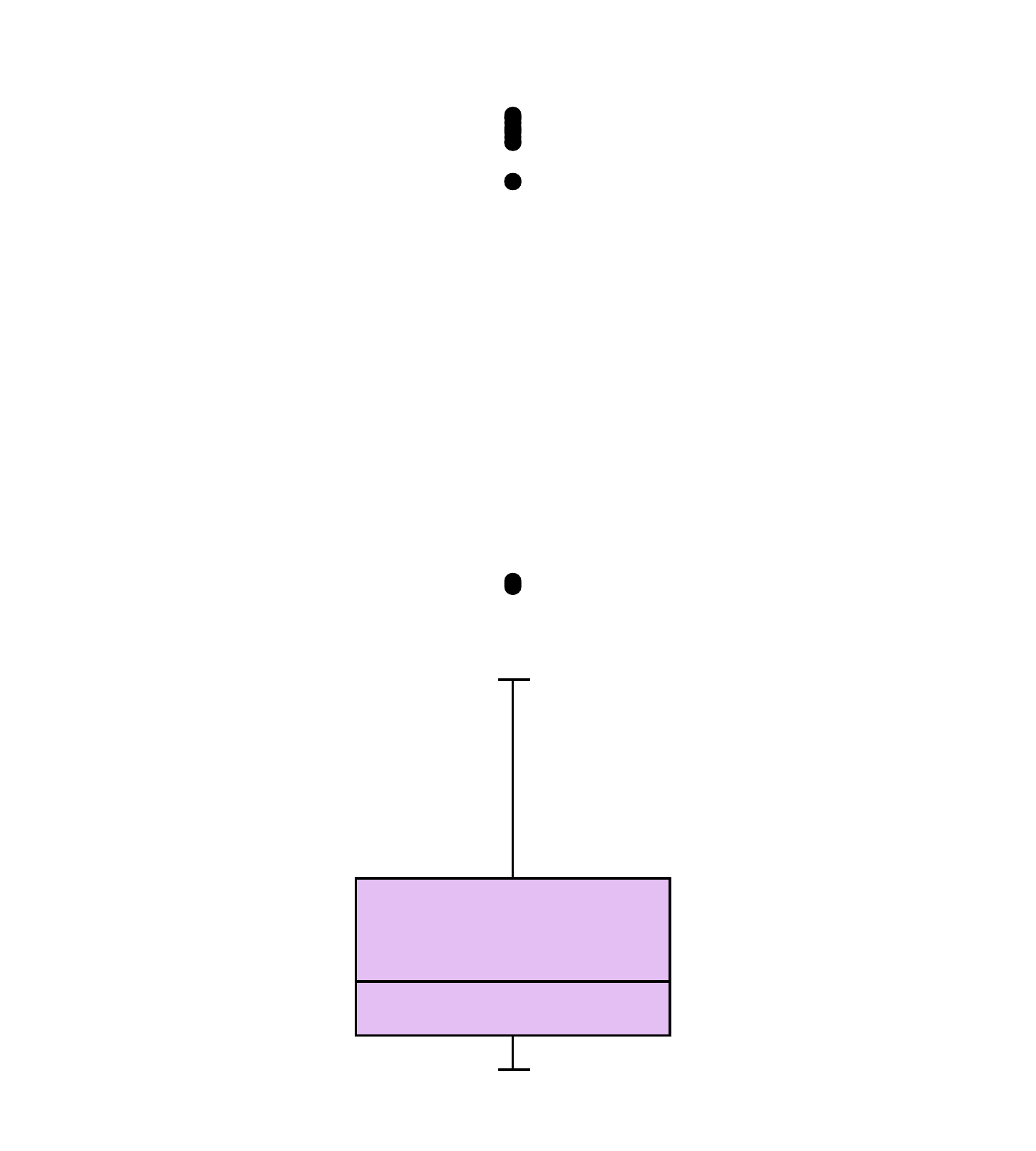}
		\label{fig:TxConfirmationDurations_Async_Boxplot}}
	\caption{Box Plots Illustrating the Distribution of Confirmation Duration [min]}
\end{figure}

Hence, we analyzed the process transaction confirmation time in greater detail. The distribution of all non-greedy transaction confirmation waiting durations is illustrated as a box plot in Figure~\ref{fig:TxConfirmationDurations_Sync_Boxplot}. A median transaction confirmation time of $7.74$~minutes was recorded for all process transactions in the non-greedy mode. This is even slightly faster than the median confirmation time of $10$ minutes the Bitcoin network is configured to exhibit. Still, a lot of outliers were recorded. One particular transaction took $172.78$~minutes to confirm. This result is not surprising, given the fact that the Bitcoin block creation duration is distributed approximately exponentially \cite{franco14}. 

\paragraph{Greedy Mode Results}
\begin{figure}[th]
	\centering
	\includegraphics[width=0.9\columnwidth]{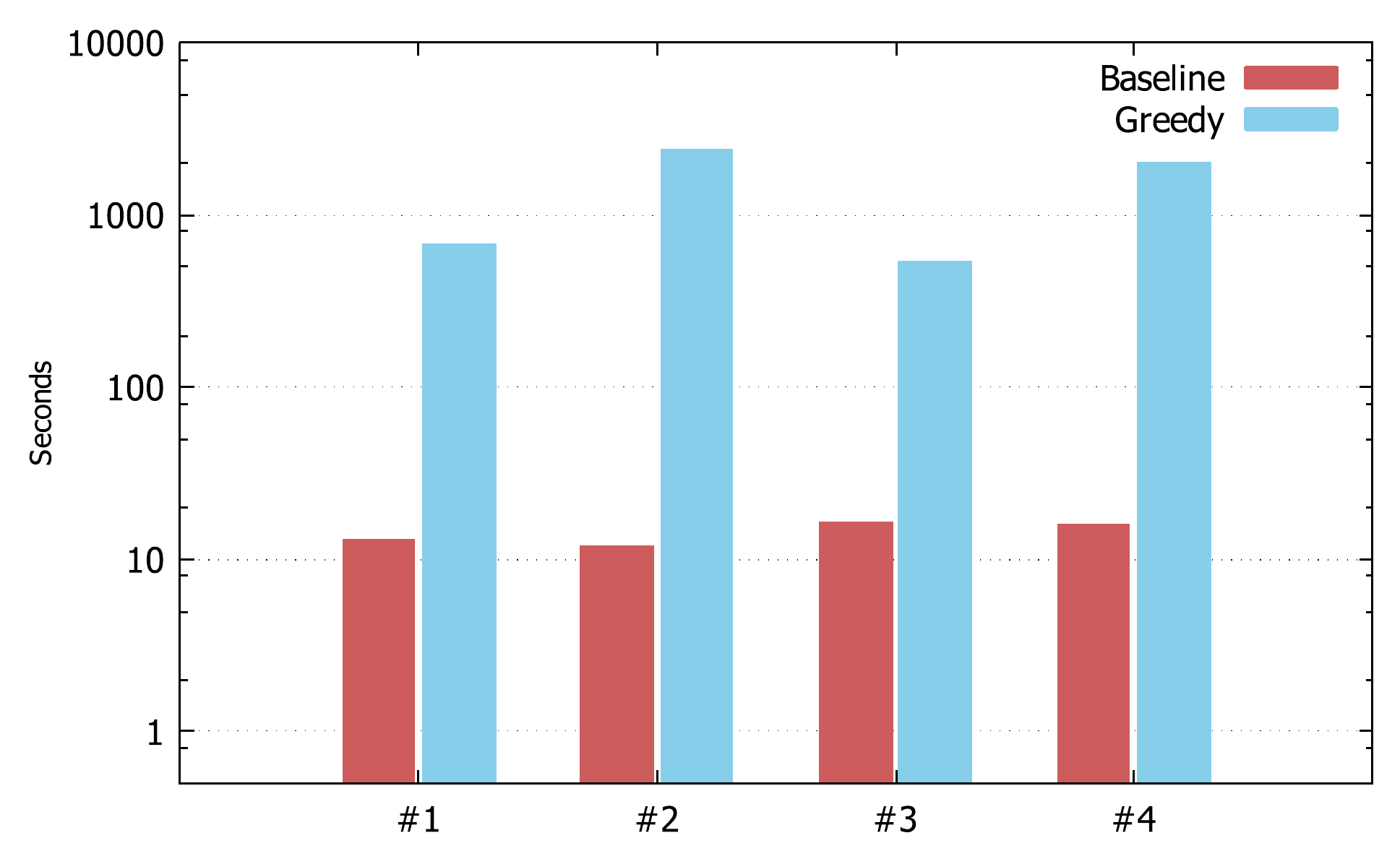}
	\caption{Bar Chart Comparing Durations of the Baseline and the Greedy Evaluation Runs}
	\label{fig:DurationComparison_Async_BarPlot}
\end{figure}
As can be seen in Figure~\ref{fig:DurationComparison_Async_BarPlot} and Table~\ref{tab:verificationFrameworkRunsGreedy}, when operating in greedy mode, the runtime verification framework still exhibits a significant increase in execution duration of the executed processes in comparison to the verification-less baseline. However, the impact of the verification framework on the execution duration is somewhat reduced in comparison the non-greedy evaluation runs. 

Again, the only difference between the baseline process execution and the greedy runtime verification execution is the conducted documentation of the single process tasks steps in the Bitcoin blockchain. Analogue to the non-greedy mode, we have analyzed which activities lead to the increase. For the greedy mode, the transaction broadcast takes 42.03\% of the overall duration, while the framework logic takes 12.47\% and the REST API requests take 45.50\%. This is a much more diverse result than discussed above for the non-greedy mode. The REST API requests and the Bitcoin transaction broadcasts have roughly the same impact to the duration of a single process task. Furthermore, the mean duration of $12.3$~seconds is much lower than the $879.68$~seconds of the non-greedy mode. The framework can potentially be even faster when being operated on top of a full Bitcoin node instead of SPV wallet. This would remove the need for the REST API requests. Most importantly, the process instances and their included tasks could be executed faster.

However, the collective waiting time for the transaction confirmations also has to be taken into account. A process instance is only considered finished when its submitted transactions have reached at least a confirmation depth of 1. The mean execution duration of the greedy evaluation runs is $1,520.68$ seconds. This duration comprises broadcasting, REST API requests, and framework logic (8.62\%) and the single transaction confirmation waiting period (91.38\%). As we can see, even though the greedy mode was able to reduce the duration of the process instances, the transaction confirmation duration remains the greatest impact factor. 

The distribution of the transaction confirmation waiting times of greedily published transactions is illustrated as a box plot in Figure~\ref{fig:TxConfirmationDurations_Async_Boxplot}.
The greedily published process transactions exhibit a median transaction confirmation time of $8.76$ minutes. Similar to the results of the non-greedily published process transactions, a large number of outliers occurred. The recorded maximum confirmation duration is $309.18$~minutes. 

The median transaction confirmation time of the greedy mode results is slightly higher than the median transaction confirmation time of the non-greedy mode results. Our experimental data does not allow to determine if this deviation is related to the chaining of unconfirmed transactions. The deviation between the two median confirmation durations can also be explained through the (roughly) exponential distribution of Bitcoin's block creation time. 

The capability of the runtime verification framework to detect incorrect executions was also evaluated. A number of evaluation runs were configured to purposefully exhibit incorrect behavior at a random handover between choreography participants. To perform incorrect behavior, a participant tries to handover a process instance to another participant with the instructions to execute a task which does not fit the current stage of the execution. This is a valid attack scenario, since the task which is supposed to be executed by the receiving participant is documented by the handover transaction that is signed by both the sending and the receiving participant. When a receiving participant detects such an incorrect behavior during a handover, it aborts the handover process. Therefore, this participant also does not sign the process handover template proposed by the sending participant, and the sending participant is not able to publish a correct process handover transaction. When the sending participant is notified that the receiver recognized the incorrect handover, it ends the execution of the process instance by publishing an extraordinary process-end transaction. During our evaluation runs, \emph{all} incorrect executions were detected.

\subsubsection{Discussion}
\label{subsub:discussion}
In both the non-greedy and greedy choreography execution results, waiting for the published transactions to be confirmed has by far the highest impact on the performance of the proposed runtime verification framework. This factor is very unpredictable, since the Bitcoin transaction confirmation duration is distributed approximately exponentially~\cite{franco14}. 

Especially in the more secure non-greedy operation mode, the runtime verification increased the execution duration of a process. During the conducted evaluation, a median transaction confirmation time of $7.74$ minutes was recorded. This shows that the runtime verification prototype is best-suited for use cases with long-running tasks. In B2B processes with tasks that take a very long time, the duration increase generated by blockchain-based runtime verification is of relatively small significance. Exemplary real-world use cases that fit this description include logistic and supply chain processes. 

The greedy execution mode of the prototype was able to reduce the overall duration by a factor of about $56$ in comparison to the non-greedy execution mode. This improvement was achieved even though the median transaction confirmation duration of the non-greedily published transaction was lower than the median of the greedily published transactions. While the framework has to wait for the confirmation of each non-greedy transaction independently, the chained greedy transactions only have one overlapping waiting time. This increase in performance is traded against a reduced level of security, as explained in Section~\ref{sub:framework}. Nevertheless, real-world business processes that include a series of very short tasks could incorporate runtime verification that operates in the greedy mode. Examples for such business process are software-centric processes which, e.g., performing a series of calculation steps. 

Due to the high standard deviation of the transaction confirmation duration, the collected results were not suited to construct a practical prediction metric, and this data should not be used to estimate the expected increase of the execution duration of processes instances. For an applicable approach to this, we refer to~\cite{2017-Yasa-ICSA}. 

Last but not least, with regard to our evaluation setup, some limitations have been assumed to simplify the implementation and should therefore be named. First, a process must only have one start and one end node, which can be achieved by process transformation~\cite{claes12}. Second, a process must be started and ended by the process owner. Since the end-process transaction cannot be used to transfer token ownership, the token must be under the control of the process owner prior to publishing it. If the last task of a business process is not executed by the process owner, the token must be transferred to the process owner in an extraordinary handover transaction. In the evaluation, these extraordinary transactions are called \emph{filler tasks} and were not further mentioned, since filler tasks are not related to any process task and serve only as a mechanism to transfer token ownership. Due to this compromise, it is possible that evaluation runs incorporated more steps than defined in the process model.

\section{Conclusion}
\label{sec:conclusion}
In this paper, we have presented an approach to runtime verification for process instances which are executed in a decentralized, choreography-based way. To enable the necessary documentation of process state updates, we propose to use blockchain technology, which allows the establishment of a decentralized trust basis. We have shown that a first generation blockchain like Bitcoin can be utilized as the foundation for runtime verification for business processes.

Furthermore, we have performed a qualitative comparison of our solution with state of the art approaches and conducted a performance analysis to determine the resulting runtime overhead. In the qualitative comparison, we have shown that our approach increases trust by applying a trustworthy, decentralized verification trust basis, while providing a large degree of flexibility. However, our approach does not take into account confidentiality of shared data. 

Because of the measured runtime overhead and the unpredictability of transaction confirmation times in the Bitcoin blockchain, our approach is primarily suitable for long-running real-world processes, e.g., in logistics or manufacturing, where the overhead only plays a minor role. However, we have also shown that by providing a greedy variant of the runtime verification approach, the overhead can be significantly reduced. This improvement comes at the cost of a reduced level of security.

The usage of blockchain for verification purposes is in an early stage, and there are many different possible research directions which need to be considered in the future. Concretely, we will investigate the realization of fault management mechanisms by the usage of Multi-Signature redeem scripts as well as the integration of further process patterns, e.g., loops.

\section*{Acknowledgements}
This paper is supported by TU Wien research funds and by the Commission of the European Union within the CREMA H2020-RIA project (Grant agreement no. 637066).

\section*{References}

\bibliographystyle{elsarticle-num} 
\bibliography{biblio}

\end{document}